\documentclass[journal]{IEEEtran}

\usepackage{amsmath,amsfonts,amssymb}
\usepackage{booktabs}
\usepackage{cite}
\usepackage{graphicx}
\usepackage{xcolor}
\usepackage[colorlinks=true]{hyperref}

\def\paperTitle{Inter-Reflective Gaussian Splatting for Robust and Efficient Inverse Rendering}
\newcommand{\irgsppt}{IRGS++}
\newif\ifshowchanges
\showchangesfalse

\ifshowchanges
  \newcommand{\ext}[1]{\textcolor{blue}{#1}}
\else
  \newcommand{\ext}[1]{#1}
\fi

\begin{document}

\title{\paperTitle}

\author{
Chun Gu,
Xiaofei Wei,
Zixuan Zeng,
Yuxuan Yao,
Li Zhang
\thanks{
Chun Gu and Xiaofei Wei are with the School of Data Science, Fudan University, and Shanghai Innovation Institute. Zixuan Zeng and Yuxuan Yao are with the School of Data Science, Fudan University. Li Zhang is with the School of Data Science, Fudan University, and Shanghai Innovation Institute. Corresponding author: Li Zhang (lizhangfd@fudan.edu.cn).
}
}

\IEEEaftertitletext{
\vspace{-0.75\baselineskip}
\begin{center}
\refstepcounter{figure}\label{fig:teaser}
\includegraphics[width=\textwidth]{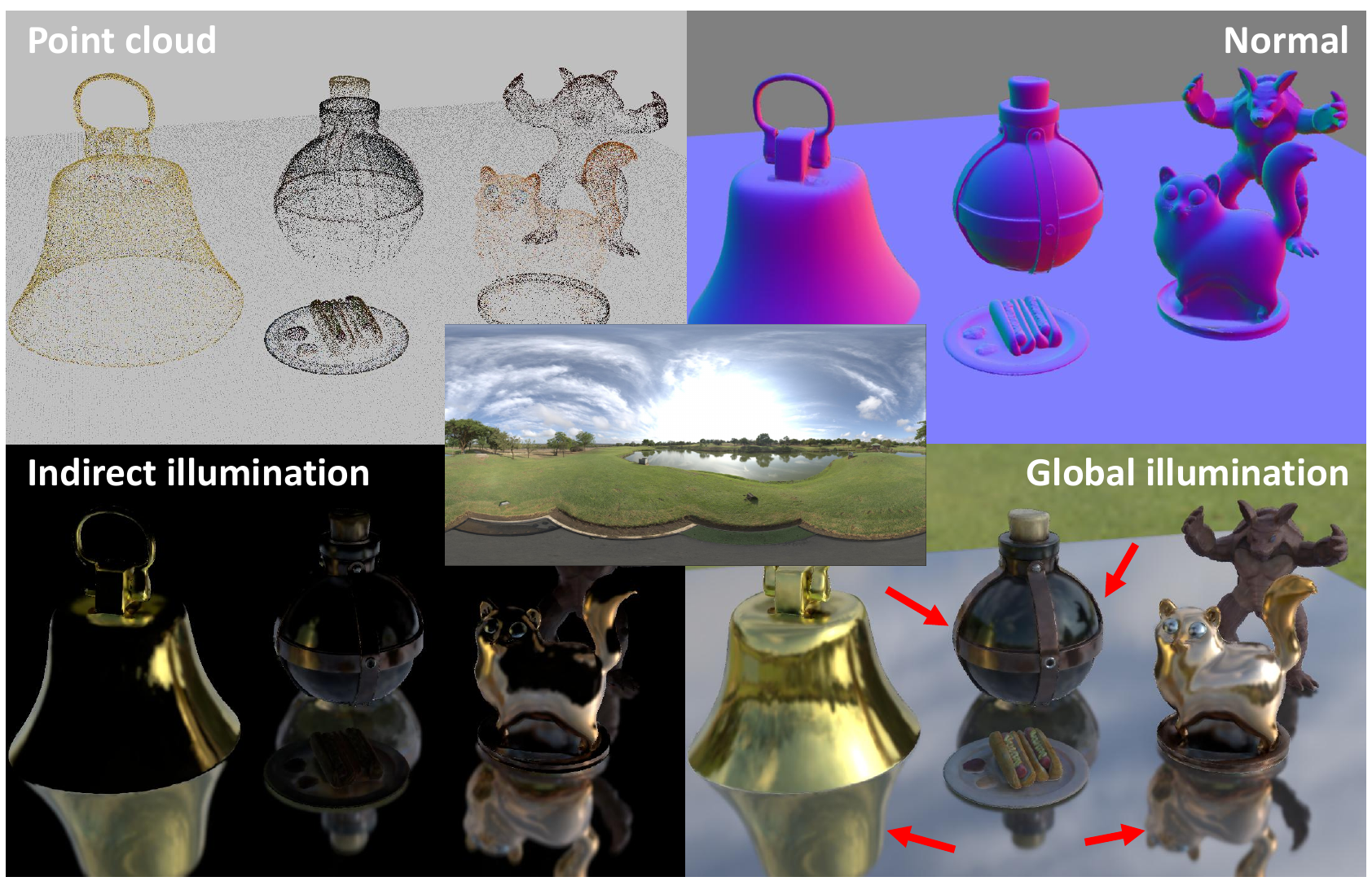}

\vspace{0.5em}
\parbox{0.98\textwidth}{\footnotesize \textbf{Fig.~\thefigure.} Point cloud, surface normals, indirect illumination, and global illumination of a relit scene composed of both low-gloss and glossy objects. The results demonstrate photorealistic secondary illumination and plausible inter-reflections with only 32 rays per pixel.}
\end{center}

\vspace{0.5\baselineskip}
}

\markboth{IEEE Transactions on Pattern Analysis and Machine Intelligence}%
{Gu \MakeLowercase{\textit{et al.}}: \paperTitle}

\maketitle

\begin{abstract}
Faithful inverse rendering requires visibility and indirect radiance to explain secondary illumination and inter-reflection, yet rasterization-oriented Gaussian representations do not naturally support the secondary-ray queries needed to recover them. We present \irgsppt{} (Inter-Reflective Gaussian Splatting), a unified robust and efficient Gaussian inverse rendering framework. During transport-aware optimization, \irgsppt{} employs differentiable 2D Gaussian ray tracing on surface-oriented Gaussian primitives to query visibility and indirect radiance on the fly and evaluate the full rendering equation for inter-reflective transport. This physical core makes Gaussian inverse rendering physically grounded beyond rasterized appearance modeling. To make this backbone useful beyond low-gloss dielectric scenes, the framework incorporates metallic-aware material modeling and robust reflective initialization for glossy, specular, and metallic materials. To make it practical, multiple importance sampling and denoising stabilize finite-sample rendering, while mesh-based secondary-attribute queries reduce the cost of relighting under novel illumination. Quantitative evaluations on low-gloss and glossy benchmarks show improved decomposition and relighting quality together with favorable quality--speed trade-offs under the reported configurations, while real-world studies illustrate plausible relighting under novel illumination.
\end{abstract}

\begin{IEEEkeywords}
Inverse rendering, Gaussian splatting, physically based rendering, relighting.
\end{IEEEkeywords}

\section{Introduction}
\label{sec:introduction}

Inverse rendering is a long-standing problem in computer vision and computer graphics. Given a set of posed images, it aims to recover scene geometry, material, and lighting, so that the reconstructed scene can be faithfully rendered and relit under novel illumination. Accurate inverse rendering is essential for relighting, material editing, and physically grounded scene understanding, all of which require the correct modeling of complex light-surface interactions.

Recent advances in scene representations have created new opportunities for inverse rendering. Neural radiance fields (NeRF)~\cite{mildenhall2021nerf} have inspired NeRF-based inverse-rendering pipelines that use repeated neural-field evaluations along rays to model visibility and incident illumination~\cite{zhang2021nerfactor,boss2021nerd,zhang2023neilf++,liu2023nero,jin2023tensoir,hasselgren2022nvdiffrecmc}. This ray-based formulation can offer high physical fidelity, but repeated neural-field queries during ray marching lead to substantial computational cost. In contrast, Gaussian splatting~\cite{kerbl3Dgaussians} provides highly efficient and high-quality scene representation, but its rasterization-based nature makes it difficult to accurately simulate ray-based effects that are central to inverse rendering.

As a result, Gaussian inverse rendering faces a three-way tension. Methods such as GS-IR and GS-ROR adopt simplified rendering equations or split-sum approximations to avoid expensive Monte Carlo integration~\cite{liang2024gs,zhu2024gs}, while R3DG uses learned or baked incident-illumination representations for efficient relighting~\cite{gao2024relightable}. These designs reduce computation, but weaken the modeling of complex secondary-light transport. Conversely, full-equation evaluation enabled by differentiable 2D Gaussian ray tracing (2DGRT) retains explicit ray-based visibility and incident-radiance queries, yet dense Monte Carlo estimation remains costly and becomes particularly high-variance for sharply peaked glossy integrands~\cite{gu2024IRGS}. Consequently, current Gaussian inverse rendering methods still struggle to jointly achieve accurate light transport, broad material coverage, and practical efficiency.

In this paper, we present \irgsppt{}, a unified robust and efficient Gaussian inverse rendering framework. Its 2DGRT formulation serves as the physical transport backbone: during transport-aware optimization, it evaluates the full rendering equation, decomposes direct and indirect illumination, and queries visibility and indirect radiance on the fly. Starting from an efficient Gaussian scene representation, \irgsppt{} performs physically based deferred rendering in image space so that geometry and material estimates are first aggregated into stable shading states before transport is evaluated. At novel-illumination relighting time, it replaces the training-time radiance query with a relighting-time secondary-attribute query backend. \ext{Around this transport backbone, the framework broadens the shading state to glossy, specular, and metallic materials, strengthens geometry and material decomposition in reflective scenes, and redesigns the relighting pipeline for practical deployment.}

\ext{Making full-equation Gaussian inverse rendering robust and practical across a broader material range requires jointly addressing material representation, estimator variance, and relighting-time query cost. First, glossy, specular, and metallic surfaces require a more expressive material model and more reliable geometry priors~\cite{yao2025refGS}; otherwise, specular highlights and reflective structures are easily misinterpreted during decomposition. Second, stratified sampling is inefficient for sharply peaked glossy integrands, making variance reduction crucial for maintaining quality at practical sample counts, as also evidenced by Monte Carlo inverse rendering methods with sampling-aware design~\cite{hasselgren2022nvdiffrecmc}. Third, although Gaussian ray tracing is effective for transport-aware optimization, directly reusing its training-time radiance queries for novel-illumination relighting is neither an efficient nor an appropriate relighting backend. To address these issues, \irgsppt{} couples metallic-aware material modeling with robust reflective-scene geometry priors, employs multiple importance sampling and denoising for low-variance rendering, and uses mesh-based ray tracing for efficient relighting.}

Quantitative experiments on synthetic low-gloss and glossy benchmarks show that the resulting framework retains the physical advantages of inter-reflective Gaussian inverse rendering while broadening material support and yielding favorable quality--speed trade-offs under the reported configurations. Qualitative studies on real-world reflective scenes further illustrate visually plausible relighting under novel illumination. Figure~\ref{fig:teaser} visualizes the point cloud, surface normals, indirect illumination, and global illumination of a relit scene composed of both low-gloss and glossy objects, illustrating physically plausible secondary lighting under practical sampling budgets.

Our contributions are summarized as follows:
\begin{itemize}
  \item We present \irgsppt{}, a Gaussian inverse rendering framework built on differentiable 2D Gaussian ray tracing, enabling full rendering-equation evaluation during transport-aware optimization with on-the-fly querying of visibility and indirect radiance for physically faithful inter-reflective rendering.
  \item \ext{We generalize physically based Gaussian inverse rendering from low-gloss dielectric scenes to glossy, specular, and metallic materials, and improve the robustness of geometry and material decomposition in highly reflective scenes.}
  \item \ext{We improve practical relighting efficiency with multiple importance sampling, denoising, and mesh-based ray tracing for relighting-time secondary-attribute queries, yielding favorable quality--speed trade-offs under the reported configurations.}
\end{itemize}

\section{Related Work}
\label{sec:related}

\paragraph{Novel View Synthesis and Scene Representation}
Novel view synthesis aims to render unseen views of a scene from a limited set of observations. Neural radiance fields (NeRF)~\cite{mildenhall2021nerf} established a powerful implicit representation for this task, and subsequent studies improved its efficiency through multiresolution encodings, voxel grids, and tensor decompositions~\cite{mueller2022instant,reiser2021kilonerf,sun2022direct,yu2021plenoctrees,Chen2022ECCV}, enhanced rendering quality through anti-aliasing formulations~\cite{barron2021mip,barron2022mip,barron2023zip}, and strengthened geometry reconstruction through neural surface models~\cite{li2023neuralangelo,wang2021neus,yariv2021volume}. More recently, 3D Gaussian splatting (3DGS)~\cite{kerbl3Dgaussians} introduced an efficient explicit scene representation with high-quality rasterization-based rendering. It has since inspired a broad family of methods for geometry reconstruction~\cite{huang20242d,Yu2024GOF}, dynamic scene modeling~\cite{yang2023real,yang2024deformable}, inverse rendering~\cite{gao2024relightable,liang2024gs}, 3D generation~\cite{tang2023dreamgaussian,yi2023gaussiandreamer}, street-scene understanding~\cite{yan2024street,chen2023periodic}, and robotics~\cite{yan2023gs,ji2024-graspsplats}. Despite these advantages, rasterization-based Gaussian representations are not naturally suited to simulating ray-based optical effects, which becomes a central bottleneck for physically grounded inverse rendering.

\paragraph{Gaussian Ray Tracing and Light Transport}
Accurate inverse rendering requires visibility reasoning, secondary-ray tracing, and indirect-light transport along sampled incident directions. Pure rasterization on Gaussian primitives is efficient for forward rendering, but it is not well suited to modeling shadows, inter-reflections, and other ray-dependent transport effects in a physically consistent manner. To address this limitation, 3D Gaussian ray tracing (3DGRT)~\cite{MonneLoccoz20243DGR} demonstrates that differentiable ray tracing can be performed directly on Gaussian scenes, enabling radiance evaluation along arbitrary rays. For surface-oriented Gaussian representations, 2D Gaussian splatting (2DGS)~\cite{huang20242d} provides geometrically more consistent primitives with well-defined ray-splat intersections. Building on these developments, our original IRGS method~\cite{gu2024IRGS} introduced 2D Gaussian ray tracing for inverse rendering and demonstrated full rendering-equation evaluation with on-the-fly visibility and indirect-radiance queries. \irgsppt{} retains this formulation as its physical transport foundation while supporting glossy and metallic materials with lower-variance rendering at practical cost.

\paragraph{NeRF-based Inverse Rendering}
Inverse rendering seeks to recover geometry, material, and illumination from multi-view images. Many NeRF-based inverse-rendering methods and closely related neural reflectance or relighting approaches~\cite{srinivasan2021nerv,boss2021nerd,yao2022neilf,zhang2023neilf++,verbin2022ref,boss2022samurai,attal2025flash,liu2023nero,jin2023tensoir,yang2023sireir,zhang2021nerfactor,wu2024neural,liang2023envidr,zhu2024multi,gu2025tensoflow} rely on ray marching and neural implicit fields to model complex light-surface interactions, including reflective appearance and indirect illumination. These methods are physically expressive, but their training and rendering pipelines remain computationally expensive. Related mesh-based physically based inverse-rendering systems further show that variance reduction is crucial in this regime. Nvdiffrec~\cite{munkberg2022extracting} uses a rasterized split-sum formulation for joint geometry, material, and lighting estimation, while its Monte Carlo successor Nvdiffrec-MC~\cite{hasselgren2022nvdiffrecmc} adds Monte Carlo rendering, learned denoising, and multiple importance sampling (MIS)~\cite{veach1995optimally} to improve efficiency and stability.

\paragraph{Gaussian-based Inverse Rendering}
Recent work has brought inverse rendering and closely related relightable or material-aware rendering to Gaussian scene representations by attaching geometry-, material-, and lighting-related attributes to individual Gaussian primitives~\cite{liang2024gs,gao2024relightable,Jiang2023GaussianShader3G,shi2023gir,wu2024deferredgs,guo2024prtgs,dihlmann2024subsurface,gu2024IRGS,yao2025refGS,zhu2024gs,lai2025glossygs,sun2025svg,chen2025gigs}. These methods efficiently leverage explicit scene representations, but they make different compromises. Some simplify the rendering equation through split-sum approximations, such as GS-IR~\cite{liang2024gs} and GS-ROR~\cite{zhu2024gs}, which limits the fidelity of material and lighting estimation under complex indirect illumination. Other systems use learned or precomputed transport representations, deferred shading, or material-aware appearance models~\cite{gao2024relightable,Jiang2023GaussianShader3G,shi2023gir,wu2024deferredgs,guo2024prtgs,dihlmann2024subsurface}, improving efficiency or flexibility without providing an on-the-fly full-equation secondary-ray transport solver. More recent methods target reflective, glossy, or spatially varying materials~\cite{yao2025refGS,lai2025glossygs,sun2025svg,chen2025gigs}, indicating the growing importance of material generality in this line of work. Taken together, existing methods still leave the joint challenge of accurate light transport, broad material coverage, and practical efficiency unresolved.

\section{Method}
\label{sec:method}

\begin{figure*}[t]
\centering
\includegraphics[width=\textwidth]{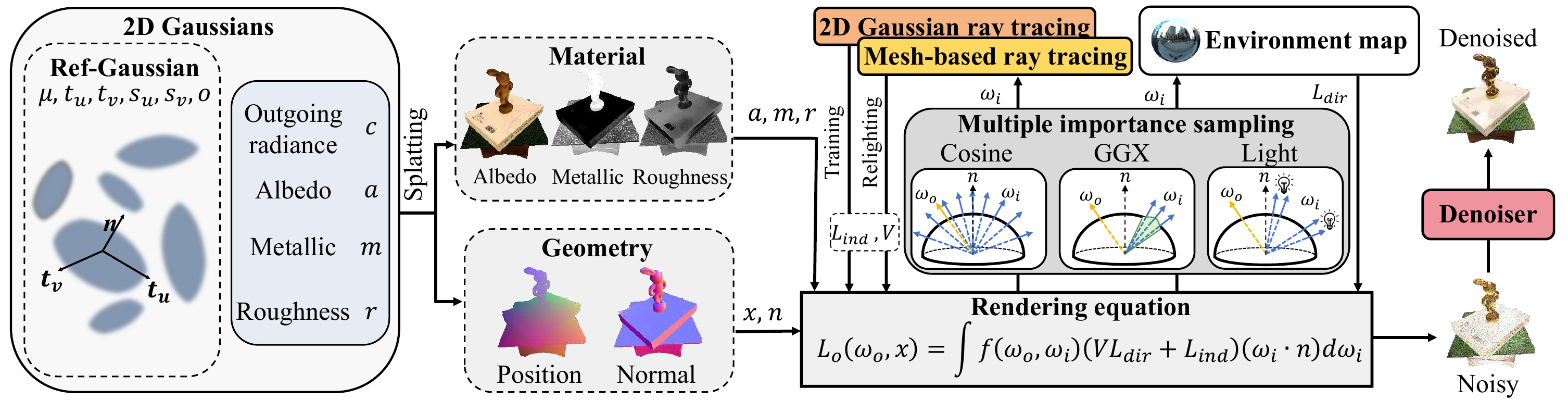}
\caption{
\textbf{Overview of \irgsppt{}.} Starting from a robust Gaussian initialization, we rasterize geometry and material attributes into pixel-space maps, evaluate the rendering equation with direct and indirect illumination, reduce Monte Carlo variance with multiple importance sampling and cross-bilateral denoising, and switch from 2D Gaussian ray tracing during training to mesh-based ray tracing during relighting for efficient secondary-light queries.
}
\label{fig:pipeline}
\end{figure*}

We present \irgsppt{}, a Gaussian inverse rendering method that combines differentiable 2D Gaussian ray tracing (2DGRT) with physically based deferred rendering. During transport-aware training, \irgsppt{} rasterizes geometry and material attributes into pixel-space maps and evaluates the rendering equation with direct and indirect illumination queried on the fly. For relighting under novel illumination, it uses a separate relighting-time secondary-query backend rather than reusing the training-time radiance trace. \ext{Within this framework, metallic-aware material modeling and robust reflective initialization support glossy and metallic scenes, while variance reduction and mesh-based ray tracing provide low-sample rendering and efficient relighting.} The pipeline follows a two-stage design: a robust first stage provides geometry and material priors, and a transport-aware second stage optimizes a 2DGRT-compatible surface representation with full rendering-equation supervision. Unless otherwise stated, all training-time light transport queries are performed on 2D Gaussian primitives, while mesh extraction is introduced only for relighting-time acceleration. Figure~\ref{fig:pipeline} summarizes the full pipeline.

\subsection{Preliminary}
\label{sec:preliminaries}

\paragraph{Gaussian splatting}
3D Gaussian splatting (3DGS)~\cite{kerbl3Dgaussians} represents a scene as a set of anisotropic Gaussian primitives. Each Gaussian is parameterized by a center $\boldsymbol{\mu}\in\mathbb{R}^{3}$, a covariance matrix $\Sigma\in\mathbb{R}^{3\times 3}$, an opacity $o\in[0,1]$, and a view-dependent outgoing radiance $\boldsymbol{c}$ modeled by spherical harmonics. The spatial influence of a Gaussian at point $\boldsymbol{x}$ is
\begin{equation}
G(\boldsymbol{x})=
\exp\left(
-\frac{1}{2}
(\boldsymbol{x}-\boldsymbol{\mu})^{\top}
\Sigma^{-1}
(\boldsymbol{x}-\boldsymbol{\mu})
\right).
\label{eq:gaussian_influence}
\end{equation}
After view transformation and perspective projection, the projected covariance can be approximated as $\Sigma' = J W \Sigma W^\top J^\top$, where $W$ denotes the camera transform and $J$ is the Jacobian of perspective projection. The rasterized color image is then obtained by front-to-back alpha blending:
\begin{equation}
\mathcal{C}=
\sum_{i=1}^{N} T_i \alpha_i \boldsymbol{c}_i,
\qquad
T_i=\prod_{j=1}^{i-1}(1-\alpha_j),
\label{eq:alpha_blending}
\end{equation}
where $\alpha_i$ combines the opacity and projected influence of the $i$-th Gaussian. In \irgsppt{}, the same compositing rule forms image-space buffers, including the rasterized outgoing-radiance image $\mathcal{C}$ used later in the deferred pipeline. For transport-aware optimization, we instantiate the scene with surface-oriented 2D Gaussian primitives so that ray tracing and splatting stay geometrically consistent while sharing the same optimized appearance and material attributes.

\paragraph{2D Gaussian splatting}
While Eq.~\ref{eq:alpha_blending} follows the standard 3DGS notation, the transport-aware stage of \irgsppt{} operates on 2D Gaussian disks inherited from 2DGS~\cite{huang20242d}. Each primitive is parameterized by a center $\boldsymbol{\mu}$, two tangential directions $\boldsymbol{t}_u,\boldsymbol{t}_v$, scales $s_u,s_v$, opacity $o$, and learnable appearance and material attributes. Compared with generic 3D Gaussians, this surface-oriented representation defines an explicit tangent plane and a well-posed local coordinate system, which are crucial for consistent ray-splat intersection and physically meaningful shading. In other words, 3DGS provides the generic rasterization formalism, while 2D Gaussian primitives provide the geometric carrier on which our transport queries are actually evaluated.

\paragraph{Rendering equation}
For our non-emissive scenes under environment lighting, inverse rendering requires evaluating the rendering equation~\cite{kajiya1986rendering} at each visible surface point $\boldsymbol{x}$:
\begin{equation}
L_o(\boldsymbol{\omega}_o,\boldsymbol{x})=
\int_{\Omega}
f(\boldsymbol{\omega}_o,\boldsymbol{\omega}_i,\boldsymbol{x})
L_{\mathrm{i}}(\boldsymbol{\omega}_i,\boldsymbol{x})
(\boldsymbol{\omega}_i\cdot\boldsymbol{n})
\mathrm{d}\boldsymbol{\omega}_i,
\label{eq:rendering_equation}
\end{equation}
where $\Omega$ denotes the upper hemisphere around $\boldsymbol{n}$, $\boldsymbol{\omega}_o$ and $\boldsymbol{\omega}_i$ denote the outgoing and incident directions, respectively, $\boldsymbol{n}$ is the unit surface normal, $f$ is the BRDF, and $L_{\mathrm{i}}$ is the incident radiance. Only directions with positive cosine contribution are considered. To cover both dielectric and glossy or metallic materials, we adopt a metallic-aware microfacet BRDF~\cite{cook1982reflectance}:
\begin{equation}
f(\boldsymbol{\omega}_o,\boldsymbol{\omega}_i,\boldsymbol{x}) = f_d + f_s,
\qquad
f_d = (1-m)\frac{\boldsymbol{a}}{\pi},
\label{eq:brdf_decompose}
\end{equation}
\begin{equation}
f_s=
\frac{
D(\boldsymbol{h},r)\,
F(\boldsymbol{\omega}_i,\boldsymbol{h},F_0)\,
G(\boldsymbol{\omega}_i,\boldsymbol{\omega}_o,\boldsymbol{n},r)
}{
4(\boldsymbol{\omega}_i\cdot\boldsymbol{n})(\boldsymbol{\omega}_o\cdot\boldsymbol{n})
},
\label{eq:brdf_specular}
\end{equation}
where $\boldsymbol{a}\in[0,1]^3$, $r\in[0,1]$, and $m\in[0,1]$ denote albedo, roughness, and metallic, respectively, $\boldsymbol{h}$ is the half vector, and $D$, $F$, and $G$ are the normal distribution, Fresnel, and geometry terms. We use the standard metallic interpolation $F_0 = 0.04(1-m)\mathbf{1} + m\boldsymbol{a}$, which recovers the dielectric case when $m=0$. Here $0.04\mathbf{1}$ denotes the channel-wise dielectric base reflectance. The half vector is defined as $\boldsymbol{h} = \frac{\boldsymbol{\omega}_i+\boldsymbol{\omega}_o}{\|\boldsymbol{\omega}_i+\boldsymbol{\omega}_o\|}$. This unified parameterization recovers dielectric materials at $m=0$ and supports glossy and metallic appearance through the joint variation of roughness and metallicity.

\paragraph{Gaussian ray tracing}
While rasterization makes Gaussian representations highly efficient, it is not naturally suited to modeling visibility, secondary rays, and inter-reflections. 3D Gaussian ray tracing (3DGRT)~\cite{MonneLoccoz20243DGR} shows that differentiable ray tracing on Gaussian primitives is feasible, and 2D Gaussian splatting~\cite{huang20242d} provides surface-oriented primitives with well-defined ray-splat intersections and more stable surface geometry. Combining these properties yielded the 2DGRT formulation introduced by IRGS~\cite{gu2024IRGS}. In \irgsppt{}, 2DGRT serves as the training-time transport operator for visibility and indirect-radiance queries on the same 2D Gaussian representation used for rasterization.

\subsection{2D Gaussian ray tracing}
\label{sec:2dgrt}

\begin{figure}[t]
\centering
\includegraphics[width=\linewidth]{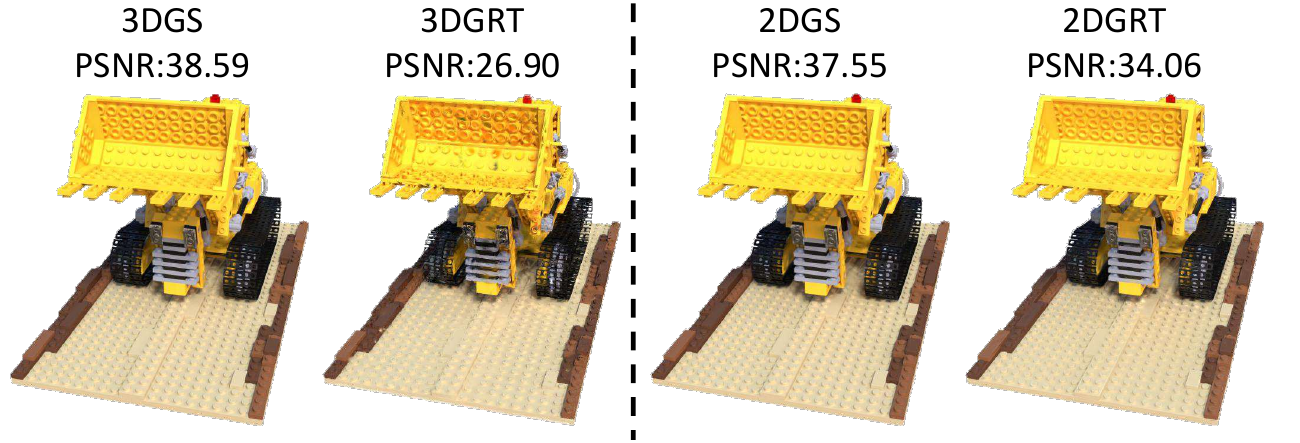}
\caption{
Ray-tracing and splatting consistency for 3D and 2D Gaussian representations. Directly tracing a pretrained 3D Gaussian splatting checkpoint causes a clear mismatch between traced and rasterized results, whereas 2D Gaussian ray tracing remains much more consistent with the splatting-based representation and is therefore better suited to inverse rendering.
}
\label{fig:raytracing-splatting}
\end{figure}

Directly applying Gaussian ray tracing to a generic 3DGS representation leads to a mismatch between ray-traced intersections and rasterized particle responses. In 3DGRT, the ray-splat interaction is defined through the maximum response of a 3D Gaussian along the ray, whereas in Gaussian splatting the particle response is accumulated in image space after projection. For inverse rendering, this inconsistency is particularly problematic because indirect illumination queries must remain as consistent as possible with the rasterized appearance and geometry supervision. Figure~\ref{fig:raytracing-splatting} illustrates that direct ray tracing on 3DGS causes a much larger degradation than ray tracing on 2D Gaussian primitives. \irgsppt{} therefore uses 2DGRT as its physical transport basis.

\paragraph{Bounding proxy and ray-splat intersection}
In the 2DGRT formulation~\cite{gu2024IRGS}, each 2D Gaussian primitive is parameterized by its center $\boldsymbol{\mu}$, opacity $o$, two principal tangential vectors $\boldsymbol{t}_u,\boldsymbol{t}_v\in\mathbb{R}^{3}$, and scales $s_u,s_v>0$. To accelerate ray tracing, we fit a conservative bounding proxy to each 2D Gaussian with an adaptive icosahedron, following the proxy construction of 3DGRT~\cite{MonneLoccoz20243DGR}. Let $\tilde{\boldsymbol{v}}$ denote a vertex of the canonical unit icosahedron. The corresponding proxy vertex is
\begin{equation}
\boldsymbol{v}
\gets
\sqrt{2\log(o/\alpha_{\min})}
\begin{pmatrix}
s_u \boldsymbol{t}_u & s_v \boldsymbol{t}_v & \epsilon \mathbf{1}
\end{pmatrix}
\tilde{\boldsymbol{v}}
+
\boldsymbol{\mu},
\label{eq:icosahedron_transform}
\end{equation}
where $\alpha_{\min}$ is the minimum enclosed influence and $\epsilon$ is a small positive constant used to obtain a valid thin proxy volume. Since each icosahedron contributes 20 faces, $N$ Gaussian disks are converted into a triangle set that can be indexed by a standard BVH. We then use hardware-accelerated OptiX ray-triangle intersection~\cite{parker2010optix} together with $k$-buffer hit sorting to recover the exact front-to-back ordering of intersected primitives.

Because 2D Gaussian disks define an explicit tangent plane, the ray-splat intersection can be computed analytically for rays with $\boldsymbol{n}^{\top}\boldsymbol{r}_d\neq 0$:
\begin{equation}
\boldsymbol{p} = \boldsymbol{r}_o + \tau \boldsymbol{r}_d,
\qquad
\tau =
\frac{
\boldsymbol{n}^{\top}(\boldsymbol{\mu}-\boldsymbol{r}_o)
}{
\boldsymbol{n}^{\top}\boldsymbol{r}_d
},
\label{eq:raytracing_intersection}
\end{equation}
where $\boldsymbol{r}_o$ and $\boldsymbol{r}_d$ are the ray origin and direction, and $\boldsymbol{n}=\frac{\boldsymbol{t}_u\times \boldsymbol{t}_v}{\|\boldsymbol{t}_u\times \boldsymbol{t}_v\|}$ is the normalized Gaussian normal. The 2D Gaussian influence at the intersection point is
\begin{equation*}
G(\boldsymbol{p})=
\exp\left(-\frac{u^2+v^2}{2}\right),
\end{equation*}
with local coordinates $u=\frac{1}{s_u}(\boldsymbol{p}-\boldsymbol{\mu})^{\top}\boldsymbol{t}_u$ and $v=\frac{1}{s_v}(\boldsymbol{p}-\boldsymbol{\mu})^{\top}\boldsymbol{t}_v$.

\paragraph{Trace operator}
Once the intersected 2D Gaussians are ordered along a ray, their outgoing radiance and opacity are accumulated by the same front-to-back compositing rule as rasterization, but now along ray-ordered hits rather than projected splats. We denote this operation by
\begin{equation}
\bigl(\boldsymbol{c}_{\mathrm{rt}}, o_{\mathrm{rt}}\bigr)
\leftarrow
\mathrm{Trace}(\boldsymbol{r}_o,\boldsymbol{r}_d),
\label{eq:trace}
\end{equation}
where $\boldsymbol{c}_{\mathrm{rt}}$ is the traced radiance and $o_{\mathrm{rt}}$ is the accumulated opacity. Since the traced radiance is still computed through differentiable alpha blending of Gaussian attributes, gradients can backpropagate through the secondary-ray queries. This property is essential for optimizing indirect illumination and visibility in inverse rendering.
\begin{equation*}
\boldsymbol{c}_{\mathrm{rt}}=
\sum_{k=1}^{K} T_k \alpha_k \boldsymbol{c}_k,
\qquad
o_{\mathrm{rt}}=
\sum_{k=1}^{K} T_k \alpha_k,
\end{equation*}
with $T_k=\prod_{\ell=1}^{k-1}(1-\alpha_\ell)$. For the training-time incident-light query used later in Eq.~\ref{eq:incident_trace}, we directly reinterpret the returned pair as
\(
L_{\mathrm{ind}}(\boldsymbol{\omega}_i,\boldsymbol{x})=\boldsymbol{c}_{\mathrm{rt}}
\)
and
\(
1-V(\boldsymbol{\omega}_i,\boldsymbol{x})=o_{\mathrm{rt}}
\)
for the ray cast from $\boldsymbol{x}$ along $\boldsymbol{\omega}_i$. A small residual discrepancy between ray tracing and 2DGS rasterization may remain because 2DGS uses tile-level sorting during splatting~\cite{gu2024IRGS}, but in practice this mismatch is much smaller than that of direct 3DGS tracing and does not alter the transport formulation.

\subsection{Rendering pipeline}
\label{sec:rendering}

\paragraph{Rasterization}
\irgsppt{} uses physically based deferred rendering~\cite{gu2024IRGS}: Gaussians are first rasterized into geometry and material maps, and shading is performed afterwards in image space. This ordering is important because direct shading on each Gaussian tends to produce unstable pixel-level normals and blurred material estimates after rasterization, a phenomenon also observed in previous Gaussian inverse rendering systems~\cite{gao2024relightable}. Specifically, at each covered pixel we rasterize
\begin{equation}
\{\mathcal{C},\mathcal{D},\mathcal{N},\mathcal{A},\mathcal{R},\mathcal{M}\}
=
\sum_{i=1}^{N}
w_i
\{\boldsymbol{c}_i,d_i,\boldsymbol{n}_i,\boldsymbol{a}_i,r_i,m_i\},
\label{eq:deferred_maps}
\end{equation}
with normalized blending weights
\begin{equation*}
w_i = \frac{T_i \alpha_i}{\sum_{j=1}^{N}T_j\alpha_j}.
\end{equation*}
Here $\mathcal{C}$ is the rasterized outgoing-radiance buffer inherited from Gaussian splatting, $\mathcal{D}$ and $\mathcal{N}$ are the depth and normal maps, and $\mathcal{A}$, $\mathcal{R}$, and $\mathcal{M}$ are the pixel-level albedo, roughness, and metallic maps. \ext{The metallic map enables the same deferred pipeline to cover both low-gloss dielectric objects and glossy or metallic surfaces.}

\paragraph{Shading state}
The deferred maps in Eq.~\ref{eq:deferred_maps} provide the full shading state required by the transport solver. Specifically, $\mathcal{D}$ determines the visible surface point $\boldsymbol{x}$ for each pixel, $\mathcal{N}$ defines the local hemisphere and cosine factor in Eq.~\ref{eq:rendering_equation}, and $\mathcal{A},\mathcal{R},\mathcal{M}$ specify the local BRDF parameters used by Eqs.~\ref{eq:brdf_decompose} and \ref{eq:brdf_specular} together with the metallic interpolation defined above. This separation is important for inverse rendering: geometry and material are first aggregated into stable pixel-aligned quantities, after which light transport is evaluated at the pixel level. Compared with per-Gaussian shading, this design greatly reduces the mismatch between material supervision, normal supervision, and the actual quantities consumed by the rendering equation.

\paragraph{Light transport}
Previous Gaussian inverse rendering methods usually lack an efficient mechanism to query visibility and incident radiance for arbitrary incident directions. As a result, they often either simplify the rendering equation through split-sum-style approximations~\cite{liang2024gs,munkberg2022extracting} or introduce learned surrogates for indirect lighting~\cite{gao2024relightable}. In contrast, 2DGRT allows us to evaluate visibility and indirect radiance on the fly after stable image-space geometry and material maps have been formed. This allows the training stage to preserve full rendering-equation supervision while keeping the advantages of Gaussian splatting and deferred shading.

\paragraph{Light parametrization}
Given the surface point $\boldsymbol{x}$ recovered from $\mathcal{D}$ and the surface normal $\boldsymbol{n}$ from $\mathcal{N}$, we decompose the incident radiance into direct and indirect terms:
\begin{equation}
L_{\mathrm{i}}(\boldsymbol{\omega}_i,\boldsymbol{x})=
V(\boldsymbol{\omega}_i,\boldsymbol{x})L_{\mathrm{dir}}(\boldsymbol{\omega}_i)
+
L_{\mathrm{ind}}(\boldsymbol{\omega}_i,\boldsymbol{x}),
\label{eq:lighting_decompose}
\end{equation}
where $L_{\mathrm{dir}}$ is modeled with an environment map. During training, the indirect component and visibility are obtained directly from 2DGRT:
\begin{equation}
\bigl(
L_{\mathrm{ind}}(\boldsymbol{\omega}_i,\boldsymbol{x}),
1-V(\boldsymbol{\omega}_i,\boldsymbol{x})
\bigr)
\leftarrow
\mathrm{Trace}(\boldsymbol{x},\boldsymbol{\omega}_i).
\label{eq:incident_trace}
\end{equation}
That is, the trace result in Eq.~\ref{eq:trace} is reused here with $\boldsymbol{c}_{\mathrm{rt}}$ interpreted as the first-bounce indirect radiance and $o_{\mathrm{rt}}$ as the accumulated occlusion along the sampled incident direction. The traced radiance in Eq.~\ref{eq:incident_trace} is formed by alpha blending the learnable outgoing radiance $\boldsymbol{c}_i$, making the entire transport query differentiable and allowing gradients to flow into indirect light estimation.

The use of $\boldsymbol{c}_i$ in Eq.~\ref{eq:incident_trace} is valid during training because $\boldsymbol{c}_i$ is optimized jointly with the current scene geometry, materials, and illumination to reproduce the observed images. Under fixed training illumination, it therefore acts as a compact outgoing-radiance field attached to the Gaussian representation, allowing a secondary ray to query the radiance leaving the first hit surface toward the current shading point. In this sense, it serves as an illumination-conditioned proxy for first-bounce indirect transport under the current training lighting, which is sufficient for transport-aware optimization. The key limitation is that this interpretation is illumination-dependent: once the environment map changes, the cached outgoing radiance is no longer physically correct, which is why relighting requires the separate attribute-tracing backbone introduced in Section~\ref{sec:mesh_relight}.

\paragraph{Rendering}
With the incident radiance defined above, the physically rendered color at a pixel is estimated by Monte Carlo integration:
\begin{equation}
\boldsymbol{c}_{\mathrm{pbr}}=
\frac{1}{N_{\mathrm{r}}}
\sum_{i=1}^{N_{\mathrm{r}}}
\frac{
f(\boldsymbol{\omega}_o,\boldsymbol{\omega}_i,\boldsymbol{x})
L_{\mathrm{i}}(\boldsymbol{\omega}_i,\boldsymbol{x})
(\boldsymbol{\omega}_i\cdot\boldsymbol{n})
}{
q(\boldsymbol{\omega}_i)
},
\label{eq:montecarlo}
\end{equation}
where $N_{\mathrm{r}}$ is the number of sampled incident directions and $q$ is their sampling density. The transport formulation remains unchanged for any valid sampling density, but its variance and computational efficiency depend strongly on how well $q$ matches the rendering integrand. Uniform or stratified hemisphere sampling wastes many samples for sharply peaked glossy lobes and concentrated environment lighting. We therefore construct $q$ with the multiple-importance-sampling scheme described next.

\subsection{\ext{Variance reduction}}
\label{sec:variance}

\ext{To reduce variance without simplifying secondary-light transport, our framework combines estimator-level multiple importance sampling with image-space denoising. MIS improves sample allocation during transport-aware optimization and relighting, whereas denoising is applied only to low-sample relighting and visualization. Neither component changes the full rendering-equation formulation established above.}

\subsubsection{\ext{Multiple Importance Sampling}}
\ext{We first define the vector-valued rendering integrand as}
\begin{equation*}
\boldsymbol{g}(\boldsymbol{\omega})=
f(\boldsymbol{\omega}_o,\boldsymbol{\omega},\boldsymbol{x})
L_{\mathrm{i}}(\boldsymbol{\omega},\boldsymbol{x})
(\boldsymbol{\omega}\cdot\boldsymbol{n}).
\end{equation*}
\ext{Following multiple importance sampling (MIS)~\cite{veach1995optimally}, we use the proposal set $\mathcal{K}=\{\mathrm{c},\mathrm{g},\mathrm{e}\}$: a cosine-weighted proposal $p_{\mathrm{c}}$ for the diffuse component, a GGX proposal $p_{\mathrm{g}}$ for the specular lobe~\cite{heitz2018sampling}, and an environment-light proposal $p_{\mathrm{e}}$ generated from the intensity distribution of the environment map~\cite{Pharr2010}. All three are normalized incident-direction PDFs with respect to solid angle on the upper hemisphere $\Omega$; $p_{\mathrm{g}}$ denotes the directional PDF after mapping a sampled GGX half vector to the incident direction, and $p_{\mathrm{e}}$ includes the solid angle represented by each environment-map sample. For each $k\in\mathcal{K}$, we draw $N_k>0$ directions $\boldsymbol{\omega}_{k,j}\sim p_k$, where $j=1,\ldots,N_k$. The total sample count and deterministic mixture density in Eq.~\ref{eq:montecarlo} are}
\begin{equation*}
N_{\mathrm{r}}=\sum_{k\in\mathcal{K}}N_k,
\qquad
q(\boldsymbol{\omega})=
\frac{1}{N_{\mathrm{r}}}
\sum_{\ell\in\mathcal{K}}
N_{\ell}p_{\ell}(\boldsymbol{\omega}).
\end{equation*}
\ext{Under the balance heuristic, the weight associated with proposal $k$ is $\beta_k(\boldsymbol{\omega})=N_kp_k(\boldsymbol{\omega})/[N_{\mathrm{r}}q(\boldsymbol{\omega})]$. Substituting this weight into the standard MIS estimator gives the following compact vector-valued form:}
\begin{equation}
\boldsymbol{c}_{\mathrm{pbr}}^{\mathrm{MIS}}
=
\frac{1}{N_{\mathrm{r}}}
\sum_{k\in\mathcal{K}}
\sum_{j=1}^{N_k}
\frac{
\boldsymbol{g}(\boldsymbol{\omega}_{k,j})
}{
q(\boldsymbol{\omega}_{k,j})
}.
\label{eq:mis}
\end{equation}
\ext{Equation~\ref{eq:mis} is exactly Eq.~\ref{eq:montecarlo} evaluated with the fixed sample allocations of the three proposals; the balance heuristic is absorbed into the mixture density $q$, so no additional normalization is required. This design matches the three dominant structures of the rendering integrand: diffuse cosine falloff, glossy specular peaks, and sparse high-intensity environment lighting. In practice, the GGX proposal is critical for sharp glossy highlights, while the environment-light proposal removes many of the artifacts caused by strong directional illumination. The same MIS construction is used across sampling budgets, with different configurations changing only the allocations $N_{\mathrm{c}}$, $N_{\mathrm{g}}$, and $N_{\mathrm{e}}$.}

\subsubsection{\ext{Denoising}}
\ext{Even with MIS, very low sample counts still produce noisy Monte Carlo estimates. Inspired by Nvdiffrec-MC~\cite{hasselgren2022nvdiffrecmc}, we therefore apply a cross-bilateral denoiser based on SVGF-style spatial filtering~\cite{schied2017spatiotemporal}. Let $\boldsymbol{c}_{\mathrm{MC}}(p)$ denote the noisy low-sample Monte Carlo color at pixel $p$, and let $\mathcal{Q}(p)$ be its spatial filter neighborhood, including $p$ itself. For a neighboring pixel $p'\in\mathcal{Q}(p)$, we define the bilateral weight}
\begin{equation*}
\begin{aligned}
B(p,p')
=\,&
\exp\left(-\frac{\|p-p'\|^2}{2\sigma_s^2}\right) \\
&\times
\exp\left(
-\frac{
|\mathcal{D}(p)-\mathcal{D}(p')|
}{
\sigma_d |\nabla \mathcal{D}(p)\cdot(p-p')|+\varepsilon_d
}
\right) \\
&\times
\max\bigl(0,\mathcal{N}(p)\cdot\mathcal{N}(p')\bigr)^{\sigma_n},
\end{aligned}
\end{equation*}
\ext{where $\nabla\mathcal{D}(p)$ is the screen-space depth gradient, $\sigma_s>0$ controls the spatial bandwidth, $\sigma_d>0$ controls depth sensitivity, $\sigma_n>0$ is the normal-similarity exponent, and $\varepsilon_d>0$ is a numerical stabilizer. We normalize $\mathcal{N}$ to unit length at every pixel before evaluating the normal similarity. The denoised color is the normalized weighted average}
\begin{equation*}
\boldsymbol{c}_{\mathrm{den}}(p)=
\frac{
\sum_{p'\in\mathcal{Q}(p)}
B(p,p')\boldsymbol{c}_{\mathrm{MC}}(p')
}{
\sum_{p'\in\mathcal{Q}(p)}B(p,p')
}.
\end{equation*}
\ext{Depth and normal guidance preserve geometric edges while removing high-frequency Monte Carlo noise. We use this denoising stage only for low-sample relighting and visualization, not as a substitute for physical transport modeling during optimization.}

\subsection{\ext{Accelerated relighting with mesh-based ray tracing}}
\label{sec:mesh_relight}

\ext{Variance reduction decreases the number of incident directions required for relighting, but two relighting-specific issues remain. First, the learned outgoing radiance attached to the Gaussian representation is conditioned on the training illumination and cannot be queried as indirect radiance after the environment changes. Second, each sampled direction still requires a secondary scene query, making ordered ray--splat evaluation the dominant relighting cost. We address these issues separately: an attribute-tracing formulation defines the information required under novel illumination, and a mesh backend answers the same query efficiently.}

\subsubsection{\ext{Relighting}}
\ext{Under novel illumination, relighting cannot reuse the training-time radiance trace in Eq.~\ref{eq:incident_trace}. Instead, for each sampled incident direction we query the visibility of the novel environment together with the material state encountered along the secondary ray:}
\begin{equation}
\bigl(V_{r},\boldsymbol{a}_{r},r_{r},m_{r},\boldsymbol{n}_{r}\bigr)
\leftarrow
\mathrm{Trace}_{\mathrm{attr}}(\boldsymbol{x},\boldsymbol{\omega}_i),
\label{eq:trace_attr}
\end{equation}
\ext{where $V_{r}\in[0,1]$ denotes visibility of the novel environment along $\boldsymbol{\omega}_i$, and $\boldsymbol{a}_{r}$, $r_{r}$, $m_{r}$, and $\boldsymbol{n}_{r}$ describe the secondary surface response returned by the query backend. Let $L_{\mathrm{dir}}^{\mathrm{new}}$ denote the novel environment lighting and let $E$ denote its prefiltered representation. We form the relighting-time incident radiance as}
\begin{equation}
\begin{aligned}
L_{\mathrm{i}}^{\mathrm{relight}}(\boldsymbol{\omega}_i,\boldsymbol{x})
=
&V_{r}L_{\mathrm{dir}}^{\mathrm{new}}(\boldsymbol{\omega}_i) \\
&+
(1-V_{r})
\mathrm{SplitSum}\bigl(\boldsymbol{a}_{r},r_{r},m_{r},\boldsymbol{n}_{r},E\bigr).
\end{aligned}
\label{eq:relight_splitsum}
\end{equation}
\ext{When the secondary ray reaches the environment, the first term evaluates its direct contribution; when it encounters scene geometry, the second term approximates the outgoing radiance of the secondary surface using its queried material attributes and the novel environment map~\cite{munkberg2022extracting}. We substitute $L_{\mathrm{i}}^{\mathrm{relight}}$ for $L_{\mathrm{i}}$ in Eq.~\ref{eq:montecarlo} and evaluate the outer integral with the MIS density defined in Section~\ref{sec:variance}. For low-sample relighting, the resulting image is subsequently processed by the denoiser described above.}
\ext{A fully recursive Monte Carlo procedure would instead trace and shade additional rays after every secondary hit. Our split-sum relighting formulation preserves an explicit secondary-surface query and the dependence on local material and novel environment lighting, while avoiding recursive multi-bounce expansion. The query interface in Eq.~\ref{eq:trace_attr} is independent of the intersection representation; the following module changes only how this query is answered.}

\subsubsection{\ext{Mesh-based ray tracing}}
\ext{After transport-aware optimization, we extract a triangle mesh once from the optimized geometry using TSDF fusion~\cite{Zhou2018}. We then bake the optimized Gaussian-derived material estimates onto the extracted surface by assigning each mesh vertex fused albedo, roughness, metallic, and normal values at the corresponding surface location, yielding a compact per-vertex attribute field for relighting.}
\ext{The attribute query in Eq.~\ref{eq:trace_attr} admits two interchangeable backends. A Gaussian backend evaluates ordered ray--splat responses, aggregates their material attributes, and sets $V_{r}$ from the accumulated opacity. The mesh backend instead performs a closest-hit ray--triangle query: a miss sets $V_{r}=1$ and exposes the novel environment, whereas a hit sets $V_{r}=0$ and recovers $\boldsymbol{a}_{r}$, $r_{r}$, $m_{r}$, and $\boldsymbol{n}_{r}$ by barycentric interpolation of the first intersected triangle. The mesh backend therefore avoids evaluating many ordered Gaussian responses for every sampled direction while leaving the incident-radiance formulation in Eq.~\ref{eq:relight_splitsum} unchanged.}
\ext{Training continues to use differentiable 2DGRT on the 2D Gaussian representation, so visibility and indirect-radiance optimization are unaffected by the relighting-time representation switch. The mesh is used only after optimization to accelerate novel-illumination queries. Because this switch changes the intersection backend rather than the training objective or relighting estimator, it provides a substantial runtime reduction with only a small measured geometry and relighting-quality cost.}

\subsection{Training scheme}
\label{sec:training}

\paragraph{Two-stage training}
Reliable geometry is essential for geometry-sensitive light transport. We therefore use a two-stage optimization strategy. The second stage is the inverse-rendering stage described above, whereas the first stage provides a geometry initialization that stabilizes the subsequent transport-aware optimization. This decomposition is particularly important in glossy scenes, where geometry, material, and reflected illumination are tightly coupled and poor initialization easily leads to incorrect local minima.

\paragraph{Initialization}
\ext{We initialize the scene with Ref-Gaussian~\cite{yao2025refGS}. This reflective initialization is substantially more robust on glossy surfaces and provides geometry and material priors before the 2DGRT-based inverse-rendering stage begins. We use the converged first-stage reconstruction to warm-start the physically based stage while keeping transport-aware optimization on the 2D Gaussian representation. Concretely, the second-stage surface-oriented Gaussian primitives receive their initial surface positions, opacity, and local orientation from the first-stage reconstruction, while the recovered albedo, roughness, and metallic estimates provide the initial material state. Ref-Gaussian thus serves only as the initialization provider; the second stage performs differentiable visibility and indirect-radiance queries with 2DGRT. This division combines a glossy-aware geometry prior with the secondary-light transport required by the unified framework.}

\paragraph{Optimization}
Evaluating the rendering equation at every pixel in every iteration is prohibitively expensive. We therefore sample only a subset $\mathcal{P}$ of pixels for Eq.~\ref{eq:montecarlo} in each iteration. Given a budget of $N_{\mathrm{rays}}$ traced rays per iteration and $N_{\mathrm{r}}$ samples per rendered pixel, we evaluate the physically based renderer on at most $\lfloor N_{\mathrm{rays}}/N_{\mathrm{r}}\rfloor$ pixels per view. This strategy preserves the quality benefit of a reasonably large per-pixel sample count while keeping optimization tractable. At the same time, the rasterized outgoing-radiance image $\mathcal{C}$ is still supervised densely, which keeps the Gaussian appearance stable and provides the learnable $\boldsymbol{c}_i$ used by the training-time 2DGRT queries. The dense reconstruction loss on $\mathcal{C}$ and the sparse physically based supervision on sampled pixels therefore play complementary roles: the former stabilizes appearance and geometry over the full image plane, while the latter constrains material decomposition and lighting through the full rendering equation at selected pixels where expensive secondary-ray queries are performed.

\paragraph{Loss function}
The second-stage optimization jointly updates the Gaussian appearance, pixel-level materials, and lighting. Its objective includes Gaussian reconstruction on the rasterized outgoing-radiance image $\mathcal{C}$, supervision of the physically rendered color $\boldsymbol{c}_{\mathrm{pbr}}$, illumination regularization, and material smoothness. Let $\Pi$ denote the full pixel grid of the current training view. We evaluate $\mathcal{L}_{\mathrm{c}}$ and the smoothness losses over $\Pi$, while the transport-dependent terms are evaluated over the sampled pixel set $\mathcal{P}$. Following the white-balance prior used in inverse rendering~\cite{liu2023nero}, the per-pixel diffuse-light statistic, denoted by $\bar{\boldsymbol{L}}_{\mathrm{d}}$, is
\begin{equation*}
\bar{\boldsymbol{L}}_{\mathrm{d}}(p)
=
\frac{1}{N_{\mathrm{r}}}
\sum_{i=1}^{N_{\mathrm{r}}}
L_{\mathrm{i}}(\boldsymbol{\omega}_i,\boldsymbol{x}(p)),
\end{equation*}
\begin{equation}
\mathcal{L}_{\mathrm{light}}
=
\frac{1}{|\mathcal{P}|}
\sum_{p\in\mathcal{P}}
\sum_c
\left|
\bar{L}_{\mathrm{d}}^{\,c}(p)
-\operatorname{mean}\bigl(\bar{\boldsymbol{L}}_{\mathrm{d}}(p)\bigr)
\right|,
\label{eq:light_reg}
\end{equation}
where $c$ indexes RGB channels and $\operatorname{mean}(\cdot)$ averages the channels of its RGB-vector argument. The edge-aware smoothness regularizer, evaluated for $X\in\{\mathcal{A},\mathcal{R},\mathcal{M}\}$, is
\begin{equation*}
\mathcal{L}_{\mathrm{s},X}
=
\frac{1}{|\Pi|}
\sum_{p\in\Pi}
\|\nabla X(p)\|
\exp\bigl(-\|\nabla \mathcal{C}_{\mathrm{gt}}(p)\|\bigr).
\end{equation*}
The physically based rendering loss is evaluated only on the sampled pixel set $\mathcal{P}$:
\begin{equation*}
\mathcal{L}_{1}^{\mathrm{pbr}}
=
\frac{1}{|\mathcal{P}|}
\sum_{p\in\mathcal{P}}
\left\|
\boldsymbol{c}_{\mathrm{pbr}}(p)-\mathcal{C}_{\mathrm{gt}}(p)
\right\|_1.
\end{equation*}
The final training objective is
\begin{equation}
\begin{aligned}
\mathcal{L}
=\,&
\mathcal{L}_{\mathrm{c}}
+ \lambda_{\mathrm{pbr}}\mathcal{L}_{1}^{\mathrm{pbr}}
+ \lambda_{\mathrm{light}}\mathcal{L}_{\mathrm{light}} \\
&+
\lambda_{\mathrm{s,a}}\mathcal{L}_{\mathrm{s},\mathcal{A}}
+ \lambda_{\mathrm{s,r}}\mathcal{L}_{\mathrm{s},\mathcal{R}}
+ \lambda_{\mathrm{s,m}}\mathcal{L}_{\mathrm{s},\mathcal{M}},
\end{aligned}
\label{eq:total_loss}
\end{equation}
where $\mathcal{L}_{\mathrm{c}}$ is the Gaussian reconstruction loss on $\mathcal{C}$ over the full image plane $\Pi$, and $\mathcal{L}_{1}^{\mathrm{pbr}}$ is the L1 loss between the physically rendered output and the ground truth image over sampled pixels. \ext{The smoothness term on $\mathcal{M}$ stabilizes metallic estimation in glossy scenes.}

\section{Experiment}
\label{sec:experiments}

\begin{figure*}[t]
  \centering
  \includegraphics[width=\linewidth]{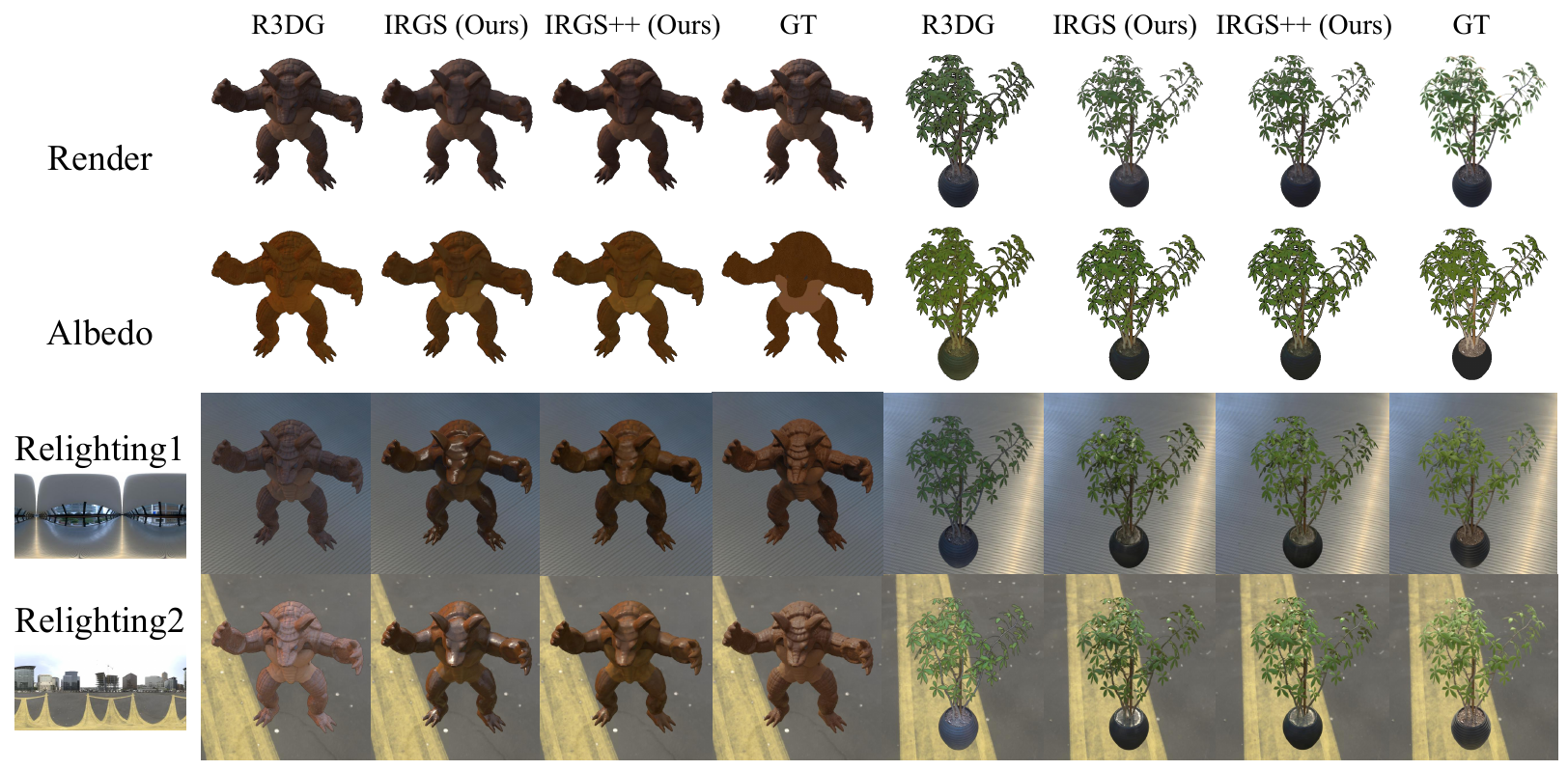}
  \caption{Qualitative comparison of NVS, albedo, and relighting results produced by R3DG~\cite{gao2024relightable}, IRGS (Ours)~\cite{gu2024IRGS}, and \irgsppt{} (Ours) on the TensoIR dataset~\cite{jin2023tensoir}.}
  \label{fig:exp_tensoir_compare}
\end{figure*}

\begin{figure*}[!t]
  \centering
  \includegraphics[width=\linewidth]{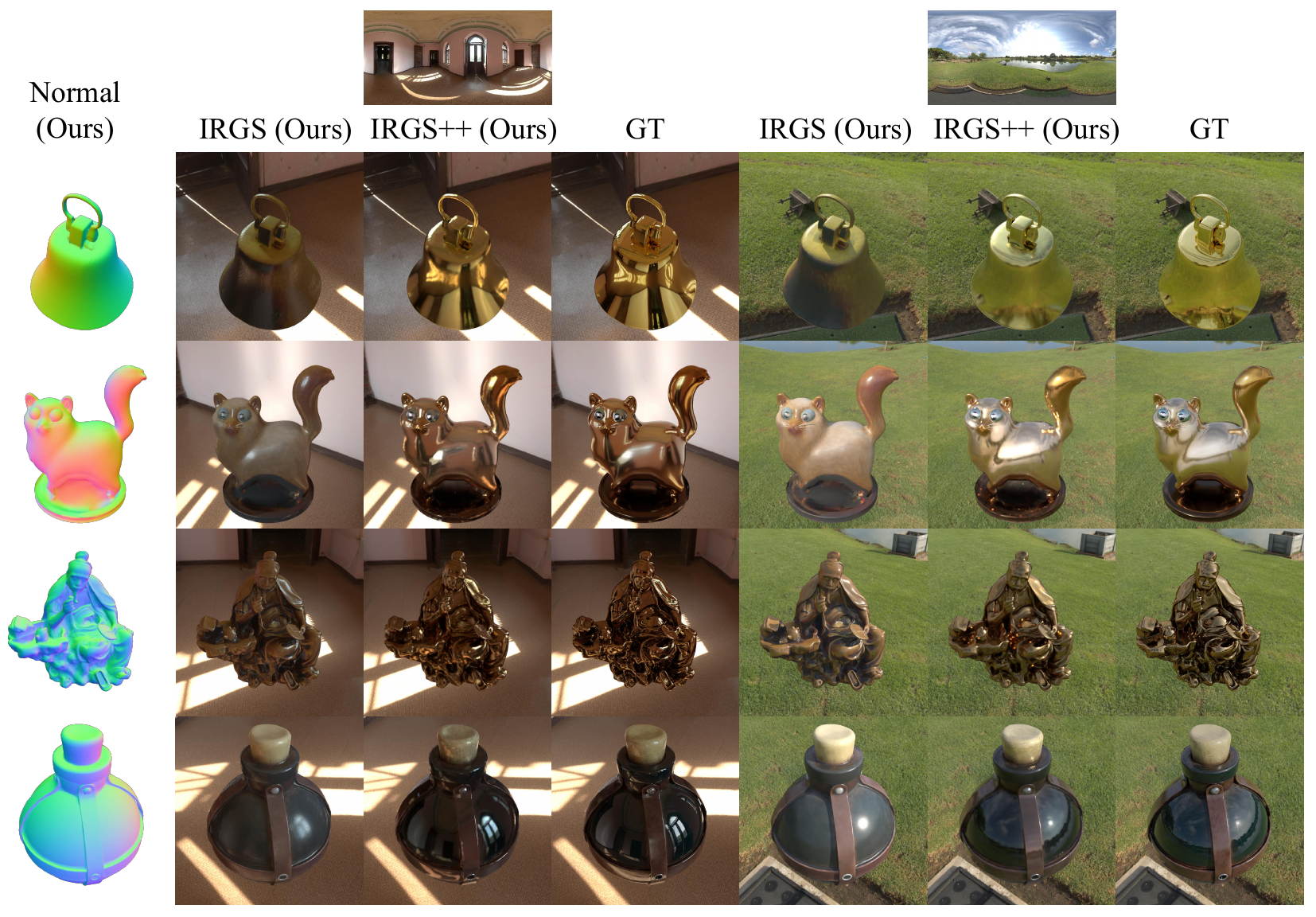}
  \caption{Relighting comparison between IRGS (Ours)~\cite{gu2024IRGS} and \irgsppt{} (Ours) on GlossySynthetic~\cite{liu2023nero} under two novel environment maps, shown above the corresponding result groups. Ground-truth relighting is shown for reference. The first column shows normal maps from the shared first-stage geometry reconstruction, which is identical for IRGS and \irgsppt{}.}
  \label{fig:exp_glossy_compare}
\end{figure*}

\subsection{Experimental settings}

\noindent\textbf{Datasets and metrics.}
We evaluate \irgsppt{} on two synthetic benchmarks and three real-world datasets. For low-gloss synthetic evaluation, we use the TensoIR dataset~\cite{jin2023tensoir} to assess normal reconstruction, novel view synthesis (NVS), albedo estimation, and relighting under both high-quality and efficient settings. For glossy and specular evaluation, we use the GlossySynthetic dataset~\cite{liu2023nero} to evaluate relighting and environment-map recovery. For real-world validation, we report qualitative relighting results on the RefReal dataset~\cite{verbin2022ref}, the GlossyReal dataset~\cite{liu2023nero}, and the Stanford-ORB dataset~\cite{kuang2024stanford}. For synthetic outputs with ground truth, we use PSNR, SSIM~\cite{wang2004image}, and LPIPS~\cite{zhang2018unreasonable} for image-valued results, and mean angular error (MAE) for surface normals. GlossySynthetic is used for metallic and highly specular scenes, while the real-world datasets are used only for qualitative evaluation because they do not provide ground-truth relit images or material maps.

\noindent\textbf{Implementation details.}
Our training procedure follows the two-stage design described in Section~\ref{sec:training}. The first stage adopts the original Ref-Gaussian configuration~\cite{yao2025refGS} to obtain robust geometry and material priors, while the second stage performs transport-aware optimization for an additional 10,000 iterations using the losses in Eq.~\ref{eq:total_loss}. We use two sampling configurations in the second stage. The high-quality (HQ) setting uses $N_{\mathrm{r}}=512$ samples per rendered pixel, allocated as 256 cosine-weighted, 128 GGX, and 128 environment-light samples. The efficient setting uses 32 samples (16/8/8) on low-gloss scenes and 16 samples (8/4/4) on glossy scenes. We implement 2DGRT and mesh-based relighting in OptiX~\cite{parker2010optix} through PyTorch CUDA extensions, and reconstruct relighting meshes with TSDF fusion~\cite{Zhou2018}. We use $64\times 128$ environment maps for low-gloss scenes and $128\times 256$ maps for glossy scenes. The complete pipeline takes about 40 minutes on a single RTX 3090 GPU, with roughly 10\,GB of memory usage.
For GlossySynthetic, we evaluate IRGS (Ours)~\cite{gu2024IRGS} with Ref-Gaussian initialization, because its original 2DGS initialization fails to reconstruct reliable geometry on highly specular scenes. This isolates the effect of transport and material modeling from failures caused purely by unstable initialization. We also keep the sampling policy fixed within each dataset rather than tuning it scene by scene. Low-gloss scenes use more rays but a lower-resolution environment map, because rough dielectric objects need lower-variance Monte Carlo integration while high-frequency lighting is less recoverable. Glossy scenes instead use fewer rays but a higher-resolution environment map, because reflective materials are more sensitive to high-frequency illumination. Unless otherwise stated, the efficient rows are reported for relighting. When runtime numbers for non-Gaussian baselines are not obtained in our own implementation stack, we report them only as references; our strongest speed comparison is between IRGS (Ours) and the two \irgsppt{} (Ours) settings under the same OptiX-based relighting pipeline.

\begin{table*}[!t]
  \centering
  \caption{Quantitative comparison of relighting quality on the GlossySynthetic dataset~\cite{liu2023nero}. Compared methods include Nvdiffrec-MC~\cite{hasselgren2022nvdiffrecmc}, TensoSDF~\cite{Li:2024:TensoSDF}, NeRO~\cite{liu2023nero}, R3DG~\cite{gao2024relightable}, GS-ROR$^2$~\cite{zhu2024gs}, Ref-Gaussian~\cite{yao2025refGS}, and IRGS (Ours)~\cite{gu2024IRGS}; the final two columns report \irgsppt{} (Ours). Training time and relighting FPS are reported from the corresponding method configurations.}
  \footnotesize
  \setlength{\tabcolsep}{1pt}
  \resizebox{\textwidth}{!}{
  \begin{tabular}{l|ccc|cccccc}
    \toprule
     & \multicolumn{3}{c|}{NeRF-based} & \multicolumn{6}{c}{3DGS-based} \\
     Scene & Nvdiff.-MC & TensoSDF & NeRO & R3DG & GS-ROR$^2$ & Ref-Gaussian & IRGS (Ours) & \irgsppt{} (Ours, 512) & \irgsppt{} (Ours, 16) \\
     & PSNR / SSIM & PSNR / SSIM & PSNR / SSIM & PSNR / SSIM & PSNR / SSIM & PSNR / SSIM & PSNR / SSIM & PSNR / SSIM & PSNR / SSIM \\
    \midrule
    Angel & \textbf{22.89} / 0.865 & 20.40 / \textbf{0.897} & 16.21 / 0.782 & 16.65 / 0.801 & 20.81 / 0.878 & 21.39 / 0.900 & 20.58 / 0.860 & \textbf{24.21} / \textbf{0.907} & 24.15 / 0.905 \\
    Bell & 24.30 / 0.903 & 29.91 / 0.977 & \textbf{31.19} / \textbf{0.979} & 16.15 / 0.839 & 24.49 / 0.927 & 22.90 / 0.920 & 20.98 / 0.878 & \textbf{25.96} / \textbf{0.934} & 25.81 / 0.932 \\
    Cat & 23.88 / 0.907 & 26.12 / 0.935 & \textbf{28.42} / \textbf{0.958} & 17.49 / 0.850 & 26.28 / \textbf{0.942} & 20.54 / 0.912 & 22.43 / 0.888 & \textbf{27.21} / 0.940 & 27.10 / 0.937 \\
    Horse & 26.42 / 0.935 & \textbf{27.18} / \textbf{0.957} & 25.56 / 0.944 & 20.63 / 0.883 & 23.31 / 0.938 & \textbf{24.97} / \textbf{0.944} & 22.10 / 0.921 & 24.81 / 0.942 & 24.80 / 0.942 \\
    Luyu & 23.60 / 0.859 & 19.91 / 0.883 & \textbf{26.22} / \textbf{0.909} & 17.47 / 0.817 & 22.61 / \textbf{0.900} & 19.74 / 0.875 & 22.73 / 0.852 & \textbf{25.74} / 0.897 & 25.73 / 0.896 \\
    Potion & 22.07 / 0.858 & 27.71 / 0.942 & \textbf{30.14} / \textbf{0.956} & 14.99 / 0.780 & 25.67 / \textbf{0.918} & 20.06 / 0.868 & 22.92 / 0.866 & \textbf{27.55} / \textbf{0.918} & 27.42 / 0.916 \\
    Tbell & 22.60 / 0.883 & 23.33 / 0.940 & \textbf{25.45} / \textbf{0.961} & 15.99 / 0.797 & \textbf{22.80} / \textbf{0.918} & 20.74 / 0.904 & 19.97 / 0.854 & 22.21 / 0.892 & 22.07 / 0.887 \\
    Teapot & 22.45 / 0.899 & 25.16 / 0.948 & \textbf{29.87} / \textbf{0.976} & 17.36 / 0.839 & 21.17 / 0.893 & 21.78 / 0.924 & 19.27 / 0.870 & \textbf{23.58} / \textbf{0.925} & 23.55 / 0.924 \\
    \midrule
    Mean & 23.53 / 0.889 & 24.97 / \textbf{0.935} & \textbf{26.63} / 0.933 & 17.09 / 0.826 & 23.39 / 0.914 & 21.51 / 0.906 & 21.37 / 0.874 & \textbf{25.16} / \textbf{0.919} & 25.08 / 0.917 \\
    \midrule
    Training Time (hours) & \textbf{4.0} & 6.0 & 12.0 & 1.0 & 1.5 & \textbf{0.6} & 1.0 & 0.7 & 0.7 \\
    Relighting FPS & \textbf{2.5} & 0.25 & 0.25 & 1.5 & 122.0 & \textbf{208.0} & 0.5 & 1.5 & 25.0 \\
    \bottomrule
  \end{tabular}
  }
  \label{tab:exp_glossy}
\end{table*}

\begin{figure}[!htbp]
  \centering
  \includegraphics[width=\linewidth]{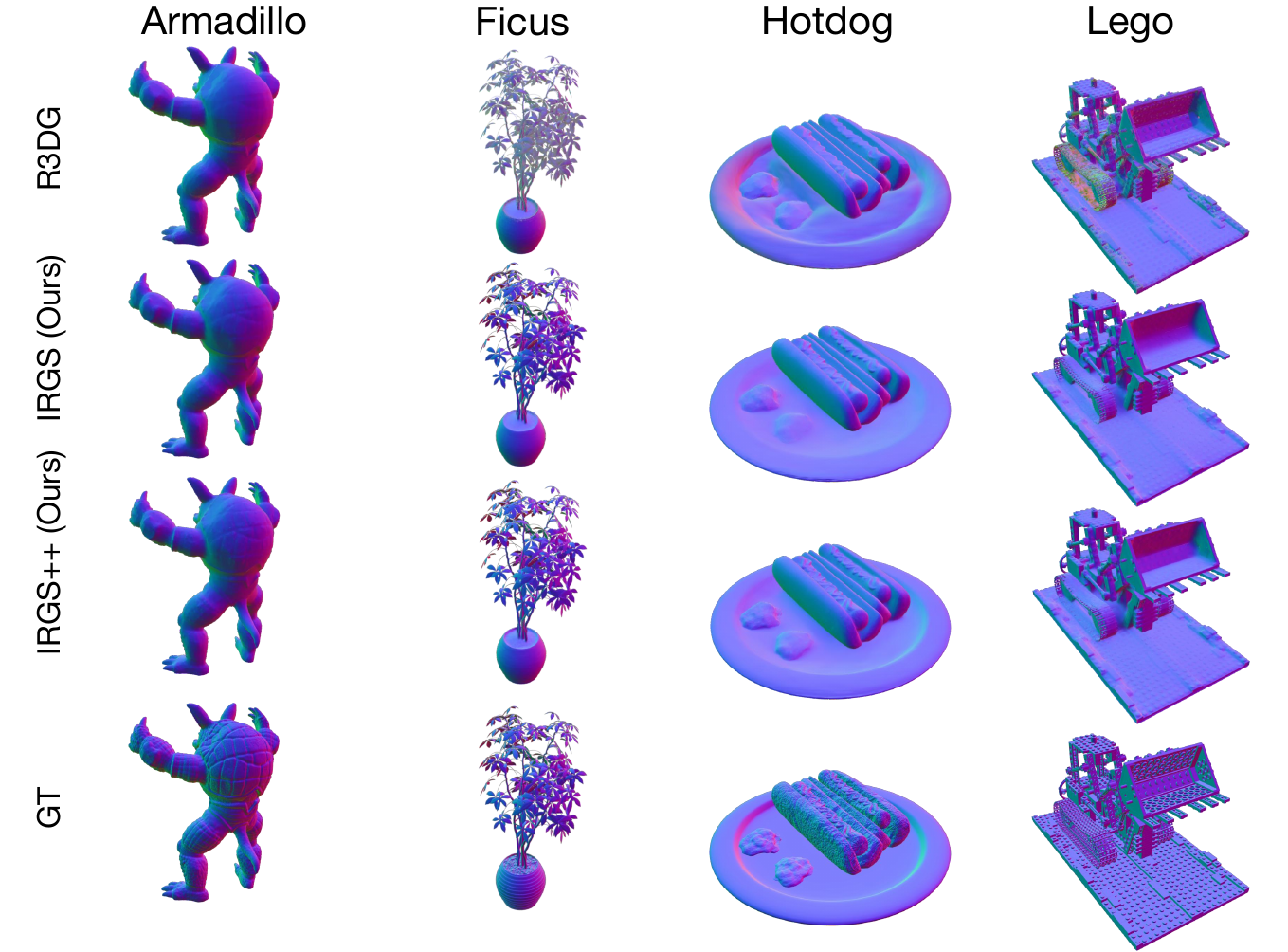}
  \caption{Rendered normal maps produced by R3DG~\cite{gao2024relightable}, IRGS (Ours)~\cite{gu2024IRGS}, and \irgsppt{} (Ours) on the TensoIR dataset~\cite{jin2023tensoir}, with ground-truth normals shown for reference.}
  \label{fig:exp_tensoir_normal}
\end{figure}

\subsection{Results on synthetic data}

\paragraph{TensoIR}
\begin{table*}[t]
  \centering
  \caption{Quantitative comparison of normal, novel-view synthesis, albedo, and relighting results on the TensoIR dataset~\cite{jin2023tensoir}. Compared methods include NeRFactor~\cite{zhang2021nerfactor}, InvRender~\cite{zhang2022modeling}, TensoIR~\cite{jin2023tensoir}, GS-IR~\cite{liang2024gs}, R3DG~\cite{gao2024relightable}, GS-ROR$^2$~\cite{zhu2024gs}, IRGS (Ours)~\cite{gu2024IRGS}, and \irgsppt{} (Ours).}
  \footnotesize
  \setlength{\tabcolsep}{3pt}
  \begin{tabular}{l|c|ccc|ccc|ccc}
    \toprule
    Method & Normal & \multicolumn{3}{c|}{Novel View Synthesis} & \multicolumn{3}{c|}{Albedo} & \multicolumn{3}{c}{Relighting} \\
     & MAE$\downarrow$ & PSNR$\uparrow$ & SSIM$\uparrow$ & LPIPS$\downarrow$ & PSNR$\uparrow$ & SSIM$\uparrow$ & LPIPS$\downarrow$ & PSNR$\uparrow$ & SSIM$\uparrow$ & LPIPS$\downarrow$ \\
    \midrule
    NeRFactor & 6.314 & 24.68 & 0.922 & 0.120 & 25.13 & 0.940 & 0.109 & 23.38 & 0.908 & 0.131 \\
    InvRender & 5.074 & 27.37 & 0.934 & 0.089 & 27.34 & 0.933 & 0.100 & 23.97 & 0.901 & 0.101 \\
    TensoIR & 4.100 & 35.09 & 0.976 & 0.040 & 29.28 & 0.950 & 0.085 & 28.58 & 0.944 & 0.081 \\
    GS-IR & 4.948 & 35.33 & 0.974 & 0.039 & 30.29 & 0.941 & 0.084 & 24.37 & 0.885 & 0.096 \\
    R3DG & 5.927 & \textbf{37.34} & \textbf{0.982} & \textbf{0.021} & 26.20 & 0.913 & 0.095 & 27.37 & 0.934 & 0.064 \\
    GS-ROR$^2$ & NA & NA & NA & NA & NA & NA & NA & 27.07 & 0.938 & 0.060 \\
    \midrule
    IRGS (Ours) & 3.998 & 35.52 & 0.964 & 0.049 & 33.42 & \textbf{0.954} & \textbf{0.076} & 30.63 & 0.935 & 0.076 \\
    \irgsppt{} (Ours, $N_{\mathrm{r}}=512$) & \textbf{3.980} & 35.15 & 0.967 & 0.045 & \textbf{33.95} & 0.949 & 0.079 & \textbf{32.12} & \textbf{0.949} & \textbf{0.059} \\
    \irgsppt{} (Ours, $N_{\mathrm{r}}=32$) & - & - & - & - & - & - & - & 31.83 & 0.944 & 0.065 \\
    \bottomrule
  \end{tabular}
  \label{tab:exp_tensoir}
\end{table*}

\begin{table}[!htbp]
  \centering
  \caption{Quantitative comparison of estimated environment maps on the GlossySynthetic dataset~\cite{liu2023nero}. Compared methods include 3DGS-DR~\cite{ye2024gsdr}, GShader~\cite{Jiang2023GaussianShader3G}, Ref-Gaussian~\cite{yao2025refGS}, IRGS (Ours)~\cite{gu2024IRGS}, and \irgsppt{} (Ours).}
  \footnotesize
  \setlength{\tabcolsep}{5pt}
  \begin{tabular}{l|ccc}
    \toprule
    Method & PSNR$\uparrow$ & SSIM$\uparrow$ & LPIPS$\downarrow$ \\
    \midrule
    3DGS-DR & 9.04 & 0.435 & 0.53 \\
    GShader & 6.52 & 0.320 & 0.61 \\
    Ref-Gaussian & 14.70 & 0.599 & \textbf{0.44} \\
    IRGS (Ours) & 7.18 & 0.304 & 0.67 \\
    \irgsppt{} (Ours) & \textbf{19.49} & \textbf{0.616} & 0.52 \\
    \bottomrule
  \end{tabular}
  \label{tab:exp_glossy_envmap}
\end{table}

\begin{figure*}[!t]
  \centering
  \includegraphics[width=\linewidth]{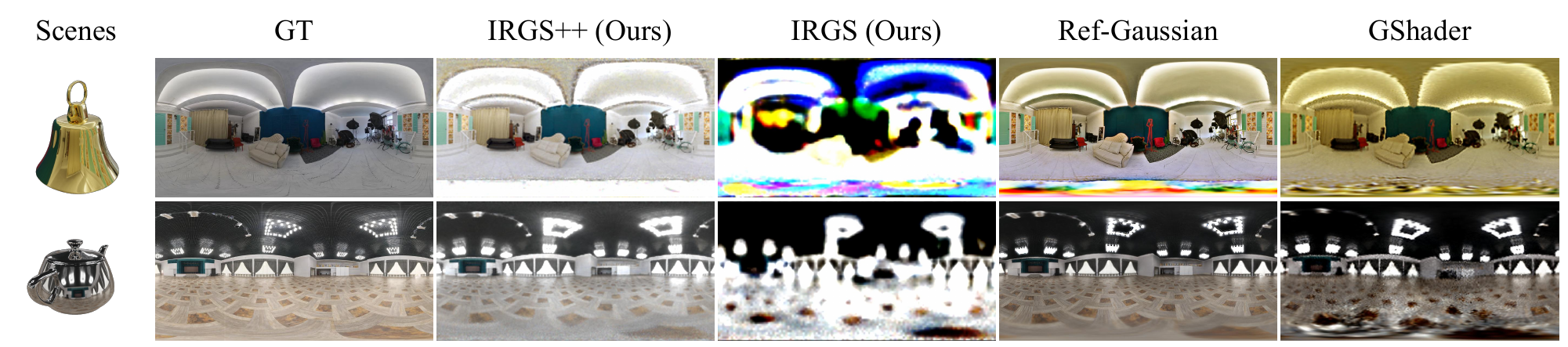}
  \caption{Estimated environment maps produced by \irgsppt{} (Ours), IRGS (Ours)~\cite{gu2024IRGS}, Ref-Gaussian~\cite{yao2025refGS}, and GShader~\cite{Jiang2023GaussianShader3G} on GlossySynthetic~\cite{liu2023nero}, with ground-truth illumination shown for reference.}
  \label{fig:exp_glossy_envmap}
\end{figure*}

\begin{figure*}[!htbp]
  \centering
  \includegraphics[width=\linewidth]{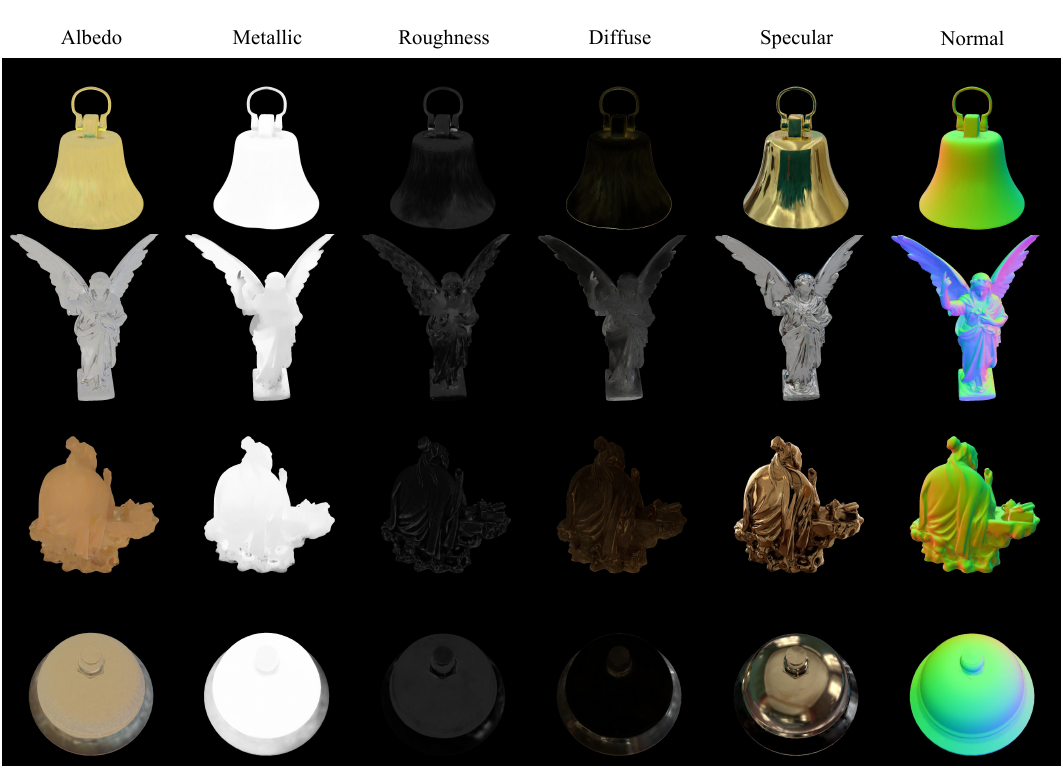}
  \caption{Albedo, metallic, roughness, diffuse, specular, and normal maps recovered by \irgsppt{} (Ours) on the GlossySynthetic dataset.}
  \label{fig:exp_glossy_material}
\end{figure*}

Table~\ref{tab:exp_tensoir} reports the main low-gloss benchmark on TensoIR. The \irgsppt{} (Ours) HQ setting achieves the best relighting quality among the compared Gaussian-based methods, while also improving albedo estimation and normal accuracy over IRGS (Ours)~\cite{gu2024IRGS}. The efficient setting with $N_{\mathrm{r}}=32$ reduces the relighting PSNR by only 0.29\,dB compared with the HQ setting, yet still remains competitive with or better than prior 3DGS-based methods. These results show that the material and efficiency improvements in \irgsppt{} increase practicality without sacrificing physically grounded decomposition on low-gloss scenes. Figure~\ref{fig:exp_tensoir_compare} further shows that \irgsppt{} produces smoother relit shadows and more stable inter-reflections, whereas IRGS (Ours) is weaker on specular cues and R3DG~\cite{gao2024relightable} misses secondary-light effects entirely.

TensoIR also isolates the quality-speed tradeoff from the glossy-specific material challenges. Since these scenes are mostly rough and dielectric, their behavior is less affected by metallic modeling and more directly reflects whether the variance-reduction strategy preserves decomposition quality. The fact that \irgsppt{} HQ improves both relighting and albedo over IRGS (Ours), while \irgsppt{} Efficient remains close to HQ, shows that the gain in efficiency is not obtained by simply shifting errors from transport to appearance.

The geometric evidence is consistent with the same conclusion. Figure~\ref{fig:exp_tensoir_normal} visualizes rendered normal maps on TensoIR and shows that the robust first stage produces cleaner surface orientation estimates than prior Gaussian baselines. Together with the normal MAE values in Table~\ref{tab:exp_tensoir}, this supports the claim that the second stage refines material and transport without destabilizing the geometry recovered in the first stage.

\begin{figure*}[t]
  \centering
  \includegraphics[width=\linewidth]{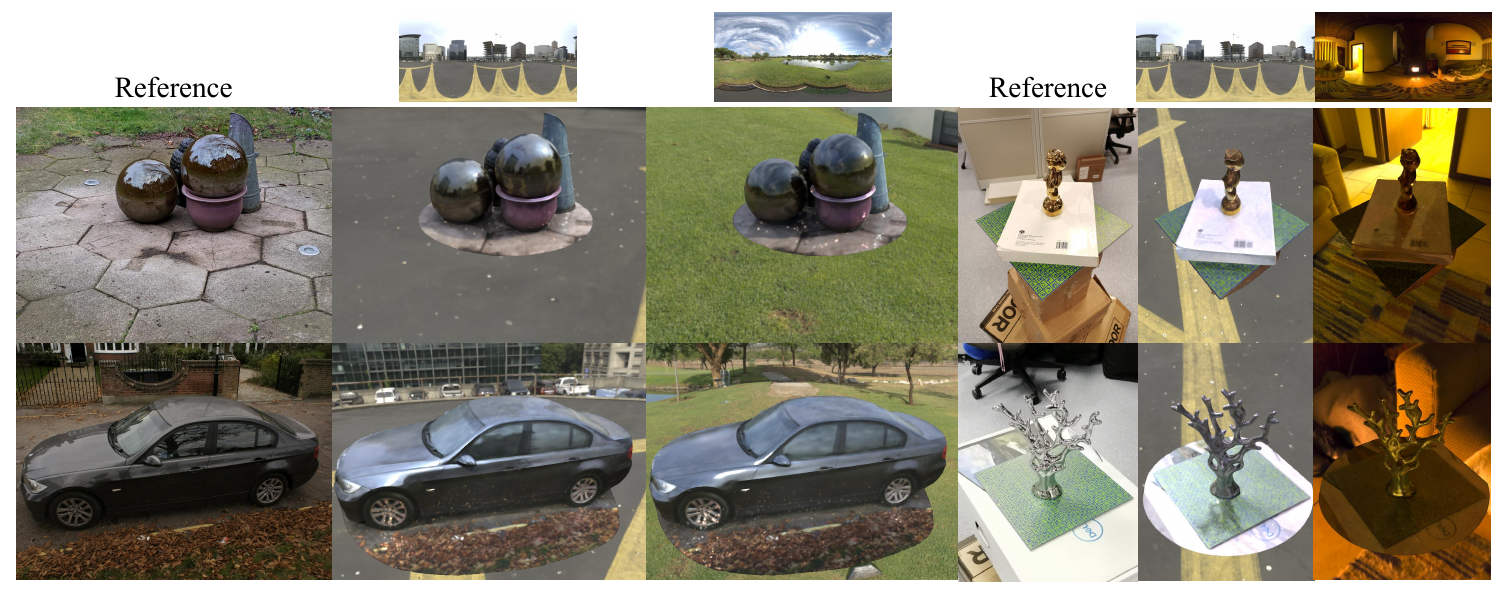}
  \caption{Relighting results on four real-world scenes from the RefReal and GlossyReal datasets. Each scene is shown in three columns: the reference input image (the ground-truth image under the captured illumination), followed by results obtained by relighting the recovered geometry and materials under two different environment maps.}
  \label{fig:exp_real}
\end{figure*}

\begin{figure*}[t]
  \centering
  \includegraphics[width=\linewidth]{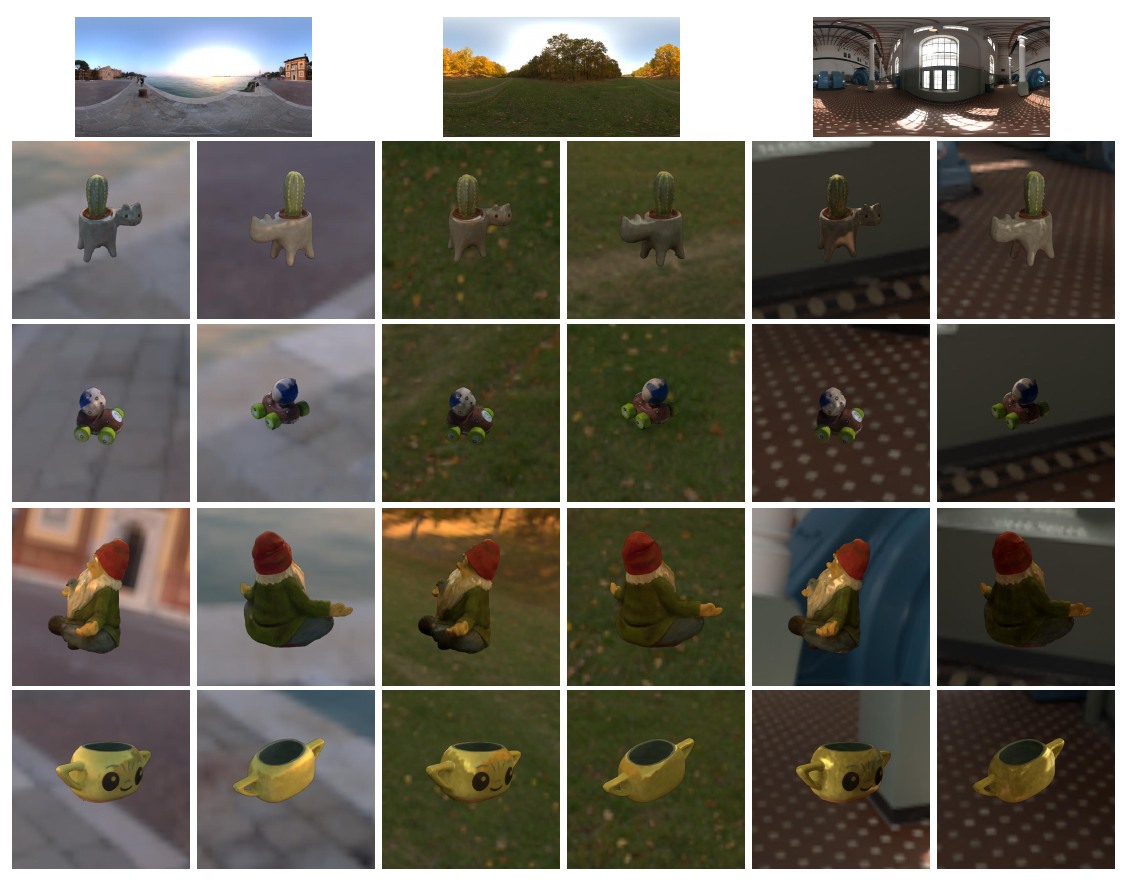}
  \caption{Relighting results on the real-world Stanford-ORB dataset. The top row shows three novel environment maps. The rows below show four reconstructed scenes relit under each environment map using their recovered geometry and materials.}
  \label{fig:exp_orb}
\end{figure*}

\paragraph{GlossySynthetic}
Table~\ref{tab:exp_glossy} reports relighting quality on the GlossySynthetic benchmark. This benchmark shows whether Gaussian inverse rendering can be extended beyond low-gloss dielectric scenes. Among Gaussian-based methods, \irgsppt{} achieves the best average relighting quality in both the HQ and efficient settings. NeRF-based methods such as TensoSDF~\cite{Li:2024:TensoSDF} and NeRO~\cite{liu2023nero} can obtain strong metrics on some scenes because they employ Blender path tracing during relighting, but they are substantially slower and are not built on Gaussian transport. For fairness, we evaluate IRGS (Ours) with Ref-Gaussian initialization on this benchmark, since its original 2DGS initialization fails to recover reliable geometry on highly specular scenes. Figure~\ref{fig:exp_glossy_compare} shows that \irgsppt{} reconstructs sharper highlight structure and more stable glossy reflections, whereas IRGS (Ours) exhibits strong artifacts due to its lack of metallic-aware modeling and low-variance glossy sampling.

The glossy benchmark is qualitatively different from low-gloss evaluation. Here the challenge is not just stronger highlights, but the combination of high metallicity, low roughness, and narrow specular lobes that make naive Monte Carlo sampling extremely inefficient. The results indicate that reliable glossy relighting depends on the combination of robust initialization, metallic-aware BRDF modeling, multiple importance sampling, and efficient secondary queries. As confirmed later in the ablation section, removing any one of these components leads to a clear degradation.

The Ref-Gaussian comparison also clarifies why it is used only for initialization. Although it yields robust geometry, it is not designed for relighting under novel illumination because its diffuse term is represented with spherical harmonics under the training lighting. The relatively low Ref-Gaussian relighting scores in Table~\ref{tab:exp_glossy} therefore reinforce our design choice: keep its geometry prior, but perform the final inverse rendering and relighting with transport-aware Gaussian primitives.

\paragraph{Environment maps}
Table~\ref{tab:exp_glossy_envmap} evaluates estimated environment maps on GlossySynthetic, and Fig.~\ref{fig:exp_glossy_envmap} provides representative visual comparisons. \irgsppt{} obtains the best PSNR and SSIM, indicating that the recovered lighting is not only visually plausible for relighting, but also quantitatively more consistent with ground-truth illumination. Although Ref-Gaussian produces visually smoother maps, its relighting and lighting decomposition remain less accurate because it models the diffuse term with spherical harmonics under the training illumination.

\begin{table*}[ht]
  \centering
  \caption{Ablation of whether 2DGRT is used during training, evaluated by relighting PSNR on GlossySynthetic.}
  \footnotesize
  \setlength{\tabcolsep}{6pt}
  \begin{tabular}{l|cccccccc|c}
    \toprule
    Setting & Angel & Bell & Cat & Horse & Luyu & Potion & Tbell & Teapot & Average \\
    \midrule
    w/ 2DGRT & \textbf{24.15} & \textbf{25.81} & \textbf{27.10} & \textbf{24.80} & \textbf{25.73} & \textbf{27.42} & 22.07 & \textbf{23.55} & \textbf{25.08} \\
    w/o 2DGRT & 23.01 & 25.17 & 25.01 & 24.19 & 25.16 & 26.56 & \textbf{22.17} & 22.76 & 24.25 \\
    \bottomrule
  \end{tabular}
  \label{tab:ablation_2dgrt}
\end{table*}

\begin{table*}[th]
  \centering
  \caption{Glossy ablation on the ``Teapot'' scene without MIS. Even 1024 stratified samples remain much worse than 16 samples with MIS.}
  \footnotesize
  \setlength{\tabcolsep}{7pt}
  \begin{tabular}{l|cccccccc}
    \toprule
    Samples & 16 & 32 & 64 & 128 & 256 & 512 & 1024 & 16 (MIS) \\
    \midrule
    PSNR$\uparrow$ & 14.59 & 14.80 & 15.20 & 15.56 & 15.87 & 16.06 & 16.15 & \textbf{23.55} \\
    \bottomrule
  \end{tabular}
  \label{tab:ablation_nomis_sweep}
\end{table*}

\begin{table}[th]
  \centering
  \caption{Normal-map MAE of rendered surfaces on TensoIR compared with GS-IR~\cite{liang2024gs}, R3DG~\cite{gao2024relightable}, and IRGS (Ours)~\cite{gu2024IRGS}. ``\irgsppt{} (Ours, mesh)'' evaluates the TSDF mesh used during \irgsppt{} relighting.}
  \footnotesize
  \setlength{\tabcolsep}{4pt}
  \resizebox{\linewidth}{!}{
  \begin{tabular}{l|ccccc}
    \toprule
    Method & GS-IR & R3DG & IRGS (Ours) & \irgsppt{} (Ours) & \irgsppt{} (Ours, mesh) \\
    \midrule
    MAE$\downarrow$ & 4.948 & 5.927 & 3.998 & \textbf{3.980} & 4.260 \\
    \bottomrule
  \end{tabular}
  }
  \label{tab:ablation_mesh_normals}
\end{table}

\begin{table}[th]
  \centering
  \caption{Efficiency-module ablations on the ``Cat'' scene from GlossySynthetic, reported in relighting PSNR.}
  \footnotesize
  \setlength{\tabcolsep}{3pt}
  \begin{tabular}{l|cccccc}
    \toprule
    Samples & 16 & 32 & 64 & 128 & 256 & 512 \\
    \midrule
    Full & \textbf{27.26} & \textbf{27.29} & \textbf{27.34} & \textbf{27.34} & \textbf{27.34} & \textbf{27.34} \\
    w/o denoiser & 25.74 & 26.08 & 26.31 & 26.40 & 26.46 & 26.48 \\
    w/o MIS & 17.83 & 18.44 & 19.38 & 20.42 & 21.49 & 22.22 \\
    w/o mesh & 27.24 & 27.26 & 27.31 & 27.31 & 27.31 & 27.31 \\
    \bottomrule
  \end{tabular}
  \label{tab:ablation_glossy_cat}
\end{table}

\begin{table}[!htbp]
  \centering
  \caption{Efficiency-oriented summary of our IRGS and \irgsppt{} configurations. Relighting PSNR is reported on TensoIR and GlossySynthetic, respectively.}
  \footnotesize
  \setlength{\tabcolsep}{3pt}
  \begin{tabular}{l|c|c|c|c|c}
    \toprule
    Setting & Rays & TensoIR & Glossy & FPS & Time (h) \\
    \midrule
    IRGS (Ours) & 512 & 30.63 & 21.37 & 0.5 & 1.0 \\
    \irgsppt{} (Ours, HQ) & 512 & \textbf{32.12} & \textbf{25.16} & 1.5 & \textbf{0.7} \\
    \irgsppt{} (Ours, Efficient) & 32 / 16 & 31.83 & 25.08 & \textbf{25.0} & \textbf{0.7} \\
    \bottomrule
  \end{tabular}
  \label{tab:efficiency_summary}
\end{table}

This result should be interpreted jointly with the relighting benchmark rather than as a standalone task. On glossy scenes, lighting estimation and material estimation are tightly coupled: an over-smoothed environment map may look natural in isolation but still lead to incorrect highlight placement and inconsistent specular transport. Our environment-map gains therefore provide additional evidence that the improved glossy relighting does not come from compensating errors in the material model, but from a more consistent recovery of both lighting and surface appearance.

We further visualize the material maps recovered by \irgsppt{} in Fig.~\ref{fig:exp_glossy_material}. Across objects with different reflective appearances, the estimated albedo, metallic, roughness, diffuse, specular, and normal components remain spatially coherent and preserve a clear separation between surface color and reflective behavior. This visualization complements the quantitative relighting results by exposing the intermediate material representation used to generalize across lighting conditions.

\subsection{Results on real-world data}
Figure~\ref{fig:exp_real} presents relighting results on RefReal and GlossyReal. Since these scenes do not provide ground-truth material maps or relit images, we focus on qualitative evaluation. The key question here is whether the recovered geometry, material, and illumination are coherent enough to support stable reflections and inter-reflections under novel lighting. In the ``gardenspheres'' example, the reflective sphere responds clearly to the changed environment while also preserving neighboring-object reflections, indicating that the glossy transport model transfers beyond synthetic data.

Stanford-ORB plays a different role, shown in Fig.~\ref{fig:exp_orb}. Unlike RefReal and GlossyReal, which emphasize scene-level reflective behavior, Stanford-ORB serves as an object-centric real-world benchmark. It therefore complements the glossy-scene evidence with a simpler setting that isolates object relighting quality. The resulting images remain visually plausible, suggesting that the same transport formulation also transfers to real captured objects under the broader material model and relighting pipeline.

Following the relighting setting used for glossy real scenes, we limit secondary-ray queries to a predefined spherical region to reduce the effect of unbounded geometry. This bounded-relighting protocol should be interpreted as a practical evaluation setting rather than a fully unconstrained real-world benchmark. Even under this restriction, the method remains challenging enough to expose unstable highlights or implausible reflections, which are largely avoided in our results.

\begin{figure*}[!t]
  \centering
  \begin{minipage}[t]{0.47\linewidth}
    \centering
    \includegraphics[width=\linewidth]{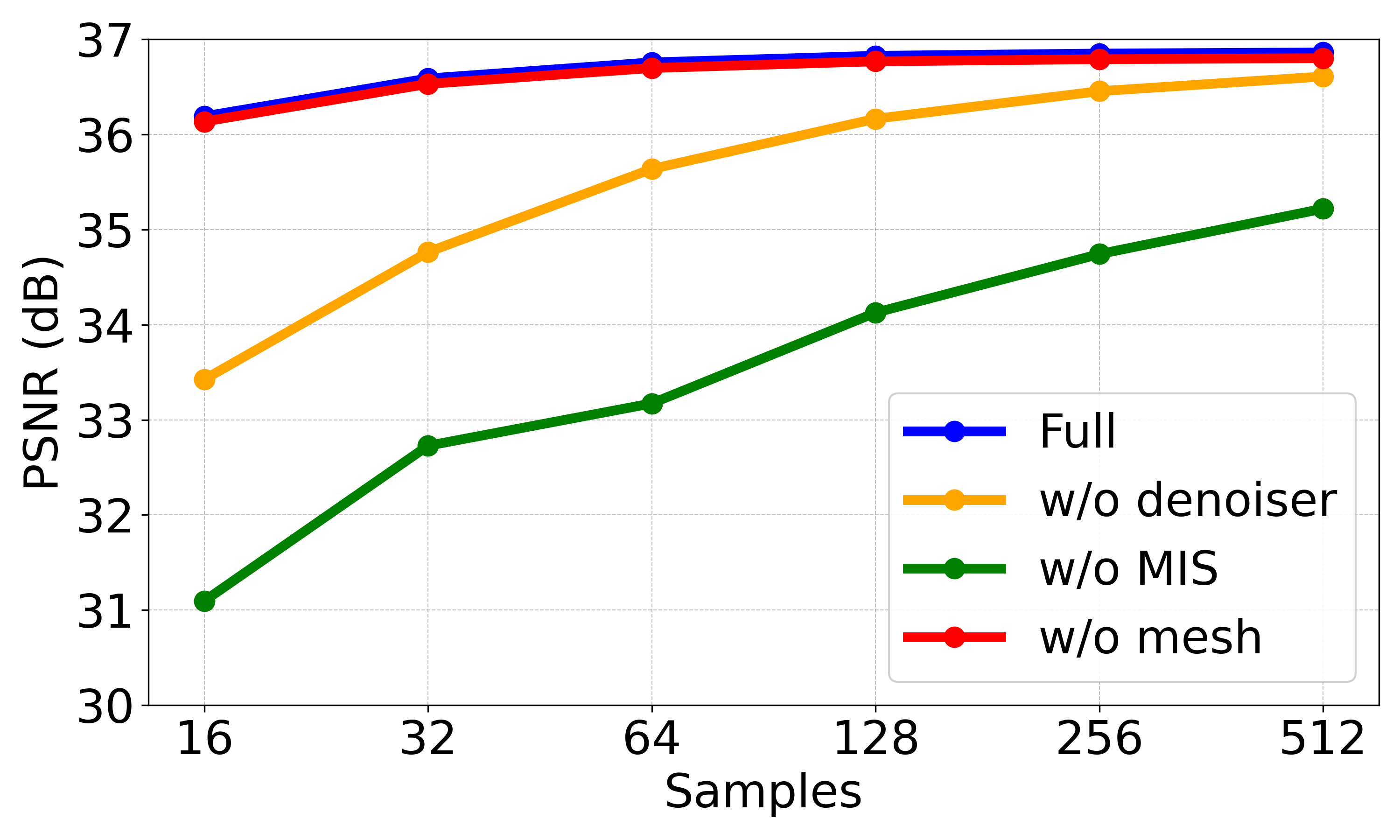}
  \end{minipage}
  \hfill
  \begin{minipage}[t]{0.47\linewidth}
    \centering
    \includegraphics[width=\linewidth]{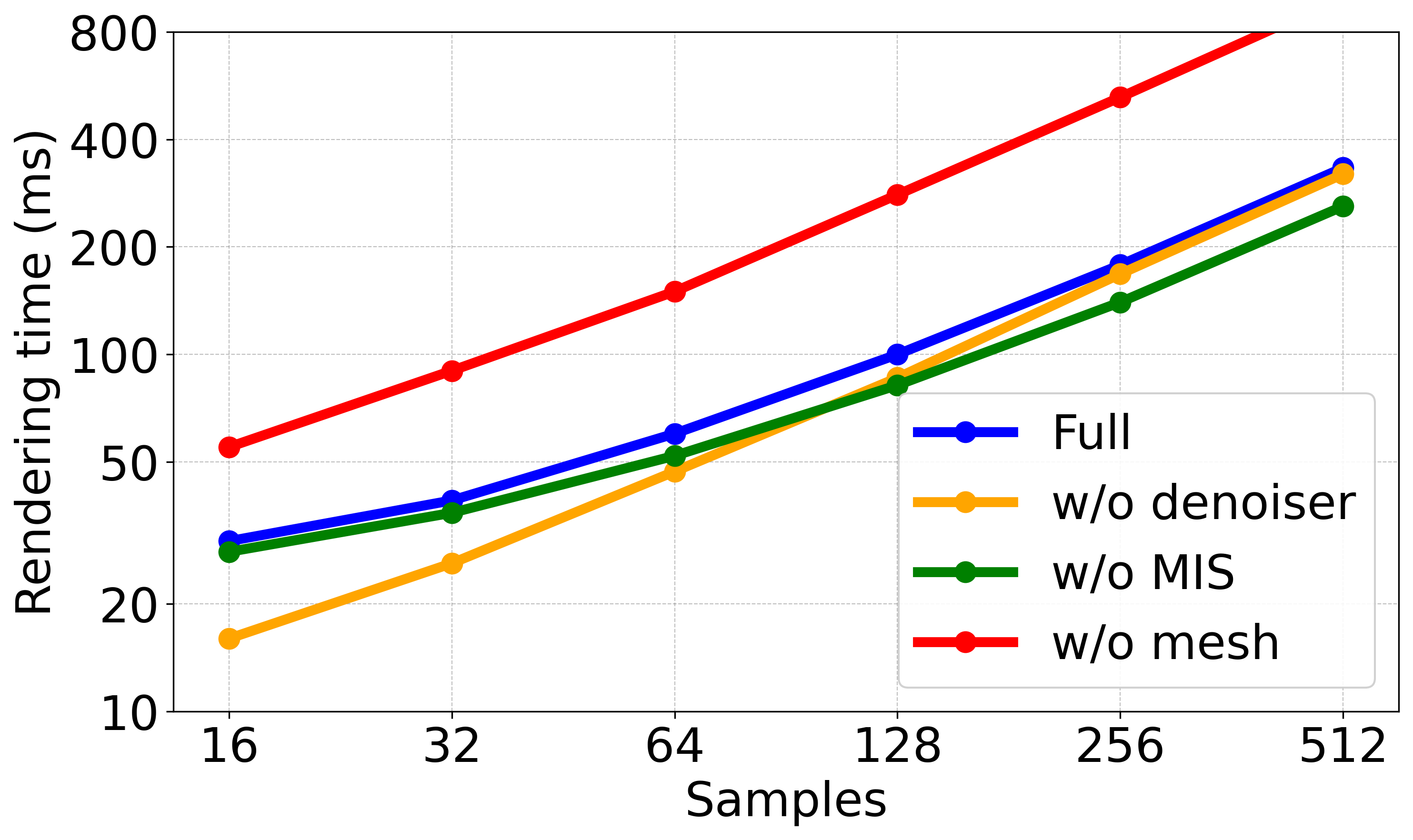}
  \end{minipage}
  \caption{Quality-speed ablations of the main efficiency modules on the ``Armadillo'' scene. Left: relighting PSNR versus sample count. Right: rendering time versus sample count.}
  \label{fig:ablation_curve}
\end{figure*}

\begin{figure*}[!t]
  \centering
  \includegraphics[width=\linewidth]{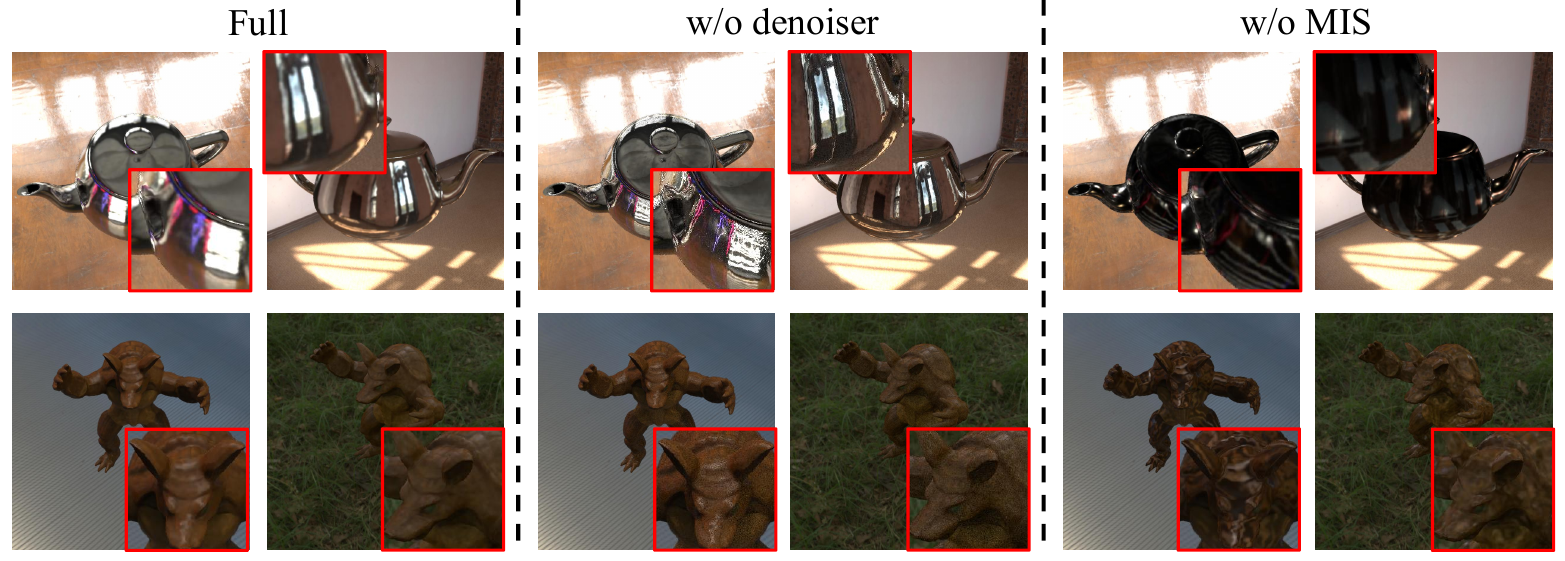}
  \caption{Visual ablations of the denoiser and multiple importance sampling. Removing the denoiser introduces Monte Carlo noise, while removing MIS degrades directional illumination and glossy highlights.}
  \label{fig:ablation_vis}
\end{figure*}

\subsection{Ablation study}

\paragraph{Training-time physical transport}
Table~\ref{tab:ablation_2dgrt} evaluates whether 2DGRT is needed during training. Replacing the differentiable transport query with a setting that does not use 2DGRT decreases the average relighting PSNR from 25.08 to 24.25 on GlossySynthetic. This validates that the training-time ray-traced transport is not a cosmetic component, but a necessary part of the final relighting quality.

\paragraph{Geometry validation}
We additionally validate two design choices that are easy to question in a two-stage pipeline. First, Table~\ref{tab:ablation_mesh_normals} compares rendered normal quality before and after converting the learned Gaussians into the TSDF mesh used for accelerated relighting. The extracted mesh incurs only a small degradation relative to the Gaussian representation and still remains better than GS-IR~\cite{liang2024gs} and R3DG~\cite{gao2024relightable}, which explains why mesh-based relighting can substantially improve speed without visibly harming quality. Second, these geometry results justify our representation split: keep the Gaussian scene representation for training-time transport, but switch to a mesh only for relighting-time intersection efficiency.

\paragraph{Efficiency}
Figure~\ref{fig:ablation_curve} studies the efficiency-oriented components of \irgsppt{} on a TensoIR scene. The full model remains stable as the sample count decreases from 512 to 16, while removing the denoiser or replacing MIS with stratified sampling causes a clear drop in low-sample quality. The ``w/o mesh'' variant stays close to the full model in PSNR, confirming that mesh-based relighting preserves quality while primarily affecting runtime. Figure~\ref{fig:ablation_vis} gives representative visual examples: without the denoiser, Monte Carlo noise is prominent; without MIS, highlights and directional illumination are estimated poorly.

These components affect different parts of the pipeline. The denoiser mainly stabilizes low-sample Monte Carlo estimates, MIS improves the probability of sampling dominant lighting directions and narrow specular lobes, and mesh-based relighting removes the dominant relighting bottleneck while leaving the rendered solution almost unchanged. The practical gains of \irgsppt{} therefore do not come from a single aggressive approximation, but from several complementary reductions of variance and runtime.

To verify that the same trend holds on glossy scenes, Table~\ref{tab:ablation_glossy_cat} reports quantitative ablations on the ``Cat'' scene from GlossySynthetic. The full model is almost unchanged from 16 to 512 samples, whereas the ``w/o MIS'' variant remains far below the full model even at 512 samples. This confirms that MIS is particularly critical once the rendering integrand becomes dominated by narrow specular lobes.

Table~\ref{tab:ablation_nomis_sweep} provides an even stricter glossy test on the ``Teapot'' scene by increasing the sample count without MIS all the way to 1024. Even this brute-force strategy remains dramatically worse than using only 16 samples with MIS, showing that the benefit of our sampling scheme is not merely faster convergence to the same estimate. In glossy scenes, the proposal distribution itself matters because the dominant specular contribution occupies too little solid angle for plain stratified sampling to recover reliably.

\subsection{Runtime}
Table~\ref{tab:efficiency_summary} summarizes the main practical configurations. On TensoIR, the efficient setting uses only $N_{\mathrm{r}}=32$ rays while losing just 0.29\,dB in relighting PSNR relative to the HQ setting. On GlossySynthetic, the efficient setting uses $N_{\mathrm{r}}=16$ rays and remains within 0.08\,dB of the 512-sample setting. This shows that our quality-speed tradeoff is already useful in both material regimes rather than being tied to a single benchmark.

The source of the speedup becomes clear when Table~\ref{tab:efficiency_summary} is read together with Fig.~\ref{fig:ablation_curve}. MIS reduces the number of wasted samples, especially on glossy scenes; the denoiser stabilizes low-sample relighting; and mesh-based ray tracing removes the dominant relighting bottleneck by replacing many ray-splat evaluations with a single ray-triangle intersection. Importantly, these gains are additive: the full model is both faster and more stable than the individual partial variants.

Compared with IRGS (Ours)~\cite{gu2024IRGS}, \irgsppt{} improves practicality along two complementary axes. The HQ setting preserves the high-fidelity transport behavior of the full rendering-equation pipeline while already improving glossy relighting quality, and the efficient setting turns this quality into a practical relighting regime. On GlossySynthetic, \irgsppt{} Efficient reaches 25 FPS, whereas IRGS (Ours) runs at 0.5 FPS under the reported configuration in Table~\ref{tab:exp_glossy}, corresponding to roughly a 50$\times$ relighting speedup. Since \irgsppt{} retains the same transport-oriented Gaussian lineage, this comparison directly reflects the practical benefit of its variance-reduction and relighting design.

\section{Conclusion}
\label{sec:conclusion}

We presented \irgsppt{}, a Gaussian inverse rendering method that combines differentiable 2D Gaussian ray tracing with physically based deferred rendering. \irgsppt{} preserves explicit rendering-equation evaluation for visibility and indirect illumination, while extending Gaussian inverse rendering from low-gloss dielectric scenes to glossy and metallic materials through a metallic-aware BRDF, robust reflective initialization, and relighting-time acceleration. As a result, the same method supports physically grounded decomposition, photorealistic relighting, and practical quality-speed tradeoffs within a single pipeline.
Extensive experiments on low-gloss, glossy, and real-world benchmarks show that the method improves relighting quality, material recovery, and efficiency over prior Gaussian-based inverse rendering approaches. The results indicate that explicit secondary-light transport, broader material modeling, and practical acceleration can be made compatible in a single Gaussian representation, rather than being treated as separate design goals.

\textbf{Limitations and Future Work.}
Despite the proposed acceleration strategies, physically based rendering-equation evaluation remains more expensive than real-time Gaussian pipelines that rely on stronger approximations. Performance also depends on reliable geometry initialization and on mesh extraction for efficient relighting, which may still introduce errors on thin structures or highly complex reflective geometry. Promising directions for future work include more efficient transport caching or learned sampling strategies, as well as geometry-aware relighting representations that preserve the speed advantage of mesh tracing without sacrificing surface fidelity.

\bibliographystyle{IEEEtran}
\bibliography{main}

% Uncomment and edit if you want author biographies.
% \begin{IEEEbiography}{Author One}
% Biography text goes here.
% \end{IEEEbiography}

\end{document}